\documentclass[11pt]{article}

\usepackage{amssymb,amsfonts,amsmath,fullpage}
\usepackage{longtable}
\usepackage{graphicx}

\begin{document}

\title{Stochastic effects in autoimmune dynamics}

\author{F. Fatehi,\hspace{0.5cm}S.N. Kyrychko, \hspace{0.5cm}A. Ross, \hspace{0.5cm}Y.N. Kyrychko, \hspace{0.5cm}K.B. Blyuss\thanks{Corresponding author. Email: k.blyuss@sussex.ac.uk}\\\\ Department of Mathematics, University of Sussex, Falmer,\\
Brighton, BN1 9QH, United Kingdom}

\maketitle

\begin{abstract}

Among various possible causes of autoimmune disease, an important role is played by infections that can result in a breakdown of immune tolerance, primarily through the mechanism of ``molecular mimicry". In this paper we propose and analyse a stochastic model of immune response to a viral infection and subsequent autoimmunity, with account for the populations of T cells with different activation thresholds, regulatory T cells, and cytokines. We show analytically and numerically how stochasticity can result in sustained oscillations around deterministically stable steady states, and we also investigate stochastic dynamics in the regime of bi-stability. These results provide a possible explanation for experimentally observed variations in the progression of autoimmune disease. Computations of the variance of stochastic fluctuations provide practically important insights into how the size of these fluctuations depends on various biological parameters, and this also gives a headway for comparison with experimental data on variation in the observed numbers of T cells and organ cells affected by infection.

\end{abstract}

\section{Introduction}

Breakdown of immune tolerance and the resulting autoimmune disease occur when the immune system fails to distinguish the host's own healthy cells from the cells affected by the infection, thus triggering an immune response that also targets healthy cells. Autoimmune disease is usually focused in a specific organ or part of the body, such as retina in the case of uveitis, central nervous system in multiple sclerosis, or pancreatic $\beta$-cells in insulin-dependent diabetes mellitus type-1~\cite{Kerr08,Prat02,San10}. Whilst it is close to impossible to pinpoint precise causes of autoimmunity in each individual case, it can usually be attributed to a number of factors, which can include the genetic predisposition, age, previous immune challenges, exposure to pathogens etc. A number of distinct mechanisms have been identified for how an infection of the host with a pathogen can result in the subsequent onset of autoimmune disease, and these include bystander activation~\cite{Fuji11} and molecular mimicry~\cite{Her96,Erco08}, which is particularly important in the context of autoimmunity caused by viral infections.

Over the years, a number of mathematical models have investigated various origins and aspects of immune response, with an emphasis on the onset and the development of autoimmune disease. Some of the earlier models studied interactions between regulatory and effector T cells without looking at causes of autoimmunity but instead focusing on T cell vaccination~\cite{Segel1995}. Borghans et al.~\cite{Borghans1995,Borghans1998} looked into this process in more detail and showed the onset of autoimmunity, which was defined as oscillations in the number of autoreactive cells that exceeded a certain threshold. Le\'on et al. \cite{Leon2000,Leon2003,Leon2004} and Carneiro et al. \cite{Carn05} have analysed interactions between different T cells and their effect on regulation of immune response and control of autoimmunity. More recently, Iwami et al. \cite{Iwami2007,Iwami2009} considered a model of immune response to a viral infection, in which they explicitly included the dynamics of a virus population. Although this model is able to demonstrate an emergence of autoimmunity, it fails to produce a regime of normal viral clearance. Alexander and Wahl \cite{Alexander2011} have focused on the role of professional antigen-presenting cells (APCs) and their interactions with regulatory and effector cells for the purposes of controlling autoimmune response. Burroughs et al. \cite{Burr11a,Burr11b} have analysed the emergence of autoimmunity through the mechanism of cytokine-mediated bystander activation. A special issue on ``Theories and modelling of autoimmunity" provides an excellent overview of recent research in the area of mathematical modelling of various aspects of onset and development of autoimmune disease~\cite{Bern15}.

These are several different frameworks for modelling the role of T cells in controlling autoimmune response. Alexander and Wahl \cite{Alexander2011} and Burroughs et al. \cite{Burr11a,Burr11b} have explicitly included a separate compartment for regulatory T cells that are activated by autoantigens and suppress the activity of autoreactive T cells. Another modelling approach is to consider the possibility of the same T cells performing different immune functions through having different or tunable activation thresholds, which allows T cells to adjust their response to T cell antigen receptor stimulation by autoantigens. This methodology was originally proposed theoretically to study peripheral and central T cell activation~\cite{GPaul92,GPaul00,Gsinger96}, and has been subsequently used to analyse differences in activation/response thresholds that are dependent on the activation state of the T cell~\cite{Bonn05}. van den Berg and Rand \cite{Berg04} and Scherer et al. \cite{Scher04} have studied stochastic tuning of activation thresholds. Interestingly, the need for T cells to have tunable activation can be shown to emerge from the fundamental principles of of signal detection theory~\cite{Noest00}. A number of murine and human experiments have confirmed that activation of T cells can indeed dynamically change during their circulation~\cite{Bit02,Nich00,Roe11,Stef02}, thus supporting the theory developed in earlier papers.

Since immune response is known to be a complex multi-factor process~\cite{Perelson1997}, a number of studies have looked into various stochastic aspects of immune dynamics, such as T cell selection and proliferation. Deenick et al. \cite{Dee03} have analysed stochastic effects of interleukin-2 (IL-2) on T cell proliferation from precursors. Blattman et al. \cite{Blat00} have shown that repertoires of the CTL (cytotoxic T cell lymphocyte) populations during primary response to a viral infection and in the memory pool are similar, thus providing further support to the theory of stochastic selection for the memory pool. Detours and Perelson \cite{Det00} have explored the distribution of possible outcomes during T cell selection with account for variable affinity between T cell receptors and MHC-peptide complexes. Chao et al. \cite{Chao04} analysed a detailed stochastic model of T cell recruitment during immune response to a viral infection. Stirk et al. \cite{Stirk2010a,Stirk2010b} have developed a stochastic model for T cell repertoire and investigated the role of competitive exclusion between different clonotypes. Using the methodology of continuous-time Markov processes, the authors computed extinction times, a limited multivariate probability distribution, as well as the size of fluctuations around the deterministic steady states. Reynolds \cite{Reyn12} have used a similar methodology to investigate an important question of asymmetric cell division and its impact on the extinction of different T cell populations and the expected lifetimes of na\"ive T cell clones. With regards to modelling autoimmune dynamics, Alexander and Wahl \cite{Alexander2011} have studied the stochastic model of immune response with an emphasis on professional APCs to show that the probability of developing a chronic autoimmune response increases with the initial exposure to self-antigen or autoreactive effector T cells. An important aspect of stochastic dynamics that has to be accounted for in the models is the so-called stochastic amplification~\cite{Alonso2007,Kuske2007}, which denotes a situation where periodic solutions with decaying amplitudes in the deterministic model can result in sustained stochastic periodic oscillations in individual realisations of the same model. This suggests that whilst on average the behaviour may show decaying-amplitude oscillations, individual realisations represented by stochastic oscillations can explain relapses/remissions in clinical manifestations of the disease as caused by endogenous stochasticity of the immune processes.

Blyuss and Nicholson \cite{Blyuss2012,Blyuss2015} have proposed and analysed a mathematical model of immune response to a viral infection that explicitly takes into account the populations of two types of T cells with different activation thresholds and also allows for infection and autoimmune response to occur in different organs. This model supports the regimes of normal viral clearance, a chronic infection, and an autoimmune state represented by exogenous oscillations in cell populations, associated with episodes of high viral production followed by long periods of quiescence. Such behaviour, that in the clinical observation could be associated with relapses and remissions, has been observed in a number of autoimmune diseases, such as MS, autoimmune thyroid disease and uveitis~\cite{Bezra95,Davies97,Nyla12}. Despite its successes, this model has a limitation that the periodic oscillations are only possible when the amount of free virus and the number of infected cells are also exhibiting oscillations, while in laboratory and clinical situations, one rather observes a situation where the initial infection is completely cleared, and this is then followed by the onset of autoimmune reaction. To overcome this limitation, Fatehi et al. (2017) have recently extended the model of Blyuss and Nicholson to also include the population of regulatory T cells and the cytokine mediating T cell activity. 

In this paper we analyse the effects of stochasticity on the dynamics of immune response in a model with the populations of T cells with different activation thresholds, regulatory cells and cytokines, as presented in Methods. Starting with a system of ordinary differential equations, we apply the methodology of continuous-time Markov chains (CTMC) to derive a Kolmogorov, or chemical master equation, describing the dynamics of a probability distribution of finding the system in a particular state. To make further analytical and numerical progress, we derive an It\^o stochastic differential equation, whose solutions provide similar stochastic paths to those of the CTMC models. This then allows us to numerically study the stationary multivariate probability distributions for the states in the model, explore stochastic amplification, determine how the magnitude of stochastic fluctuations around deterministic steady states depends on various parameters, and investigate the effects of initial conditions on the outcome in the case of bi-stability between different dynamical states. These results suggest that the experimentally observed variation in the progression of autoimmune disease can be attributed to stochastic amplification, and they also provide insights into how the variance of fluctuations depends on parameters, which can guide new laboratory experiments. 

\section{Methods}

\subsection{Continuous-time Markov chain model of immune dynamics}

In a recent paper we introduced and analysed a deterministic model for autoimmune dynamics with account for the populations of T cells with different activation thresholds and cytokines (Fatehi et al. 2017). The analysis showed that depending on parameters and initial conditions, the model can support the regimes of {\it normal disease clearance}, where an initial infection is cleared without further consequences for immune dynamics, {\it chronic infection} characterised by a persistent presence of infected cells in the body, and the state of {\it autoimmune behaviour} where after clearance of initial infection, the immune system supports stable endogenous oscillations in the number of autoreactive T cells, which can be interpreted in the clinical practice of autoimmune disease as periods of relapses and remissions. This work extended earlier results on modelling the effects of tunable activation thresholds~\cite{Blyuss2012,Blyuss2015} by including regulatory T cells, as well as the cytokine mediating proliferation and activity of different types of T cells.

\begin{figure}%[h]
    \begin{center}
    \includegraphics[width=16cm]{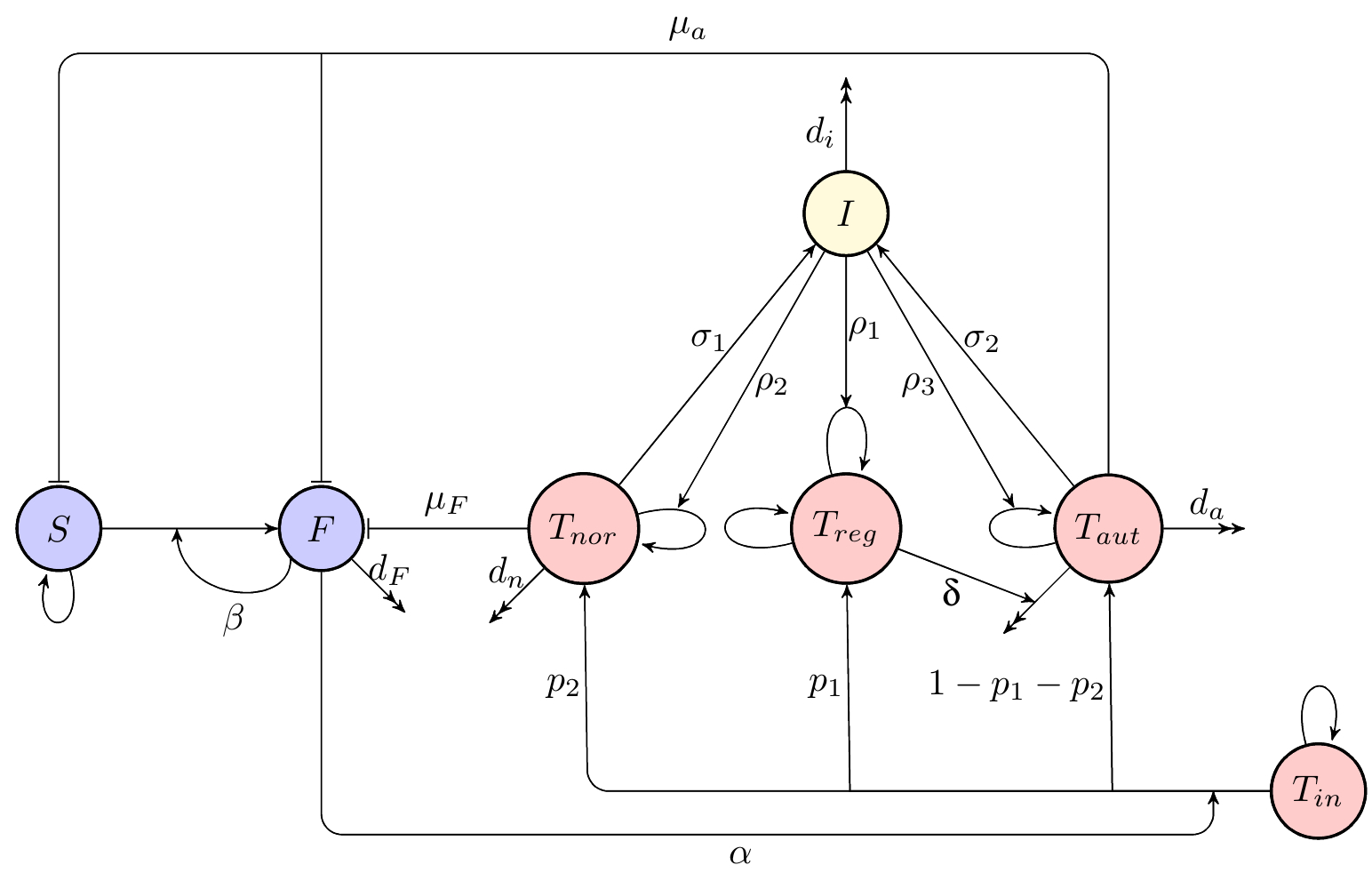}% This is a *.jpg file
    \end{center}
	\caption{A schematic diagram of immune response to infection. Blue indicates host cells (susceptible and infected), red denotes T cells (na\"ive, regulatory, normal activated, and autoreactive), yellow shows cytokines (interleukin-2).}
	\label{fig1}
\end{figure}

A deterministic model for immune response to a viral infection, as illustrated in a diagram shown in Figure~\ref{fig1},
has the form
\begin{equation}\label{firstmodel}
\begin{array}{l}
\displaystyle{\frac{dS}{dt}=rS\left(1-\frac{S}{N}\right)-\beta SF-\mu_a T_{aut}S,}\\\\
\displaystyle{\frac{dF}{dt}=\beta SF-d_FF-\mu_FT_{nor}F-\mu_a T_{aut}F,}\\\\
\displaystyle{\frac{dT_{in}}{dt}=\lambda_{in}-d_{in}T_{in}-\alpha T_{in}F,}\\\\
\displaystyle{\frac{dT_{reg}}{dt}=\lambda_r-d_rT_{reg}+p_1\alpha T_{in}F+\rho_1 IT_{reg},}\\\\
\displaystyle{\frac{dT_{nor}}{dt}=p_2\alpha T_{in}F-d_nT_{nor}+\rho_2 IT_{nor},}\\\\
\displaystyle{\frac{dT_{aut}}{dt}=(1-p_1-p_2)\alpha T_{in}F-d_aT_{aut}-\delta T_{reg}T_{aut}+\rho_3 IT_{aut},}\\\\
\displaystyle{\frac{dI}{dt}=\sigma_1T_{nor}+\sigma_2T_{aut}-d_iI,}
\end{array}
\end{equation}
where $S(t)$ is the number of susceptible organ cells, $F(t)$ is the number of infected cells, $T_{in}(t)$ is the number of na\"ive T cells, $T_{reg}(t)$ is the number of regulatory T cells, $T_{nor}(t)$ is the number of activated T cells which recognise foreign antigen and destroy infected cells, $T_{aut}(t)$ is the number of autoreactive T cells which destroy cells presenting both foreign and self-antigen, and  $I(t)$ is the amount of interleukin 2 (IL-2) cytokine. In this model, it is assumed that in the absence of infection, organ cells in the host reproduce logistically with a linear growth rate $r$ and carrying capacity $N$, and they can become infected at rate $\beta$ by already infected cells that are producing new virus particles. Unlike earlier models~\cite{Blyuss2012,Blyuss2015} and (Fatehi et al. 2017), we consider the situation where the process of producing virions by infected cells is quite fast, hence, we do not explicitly incorporate a separate compartment for free virus. Regarding immune response, we assume that na\"ive T cells remain in homeostasis, and upon activation at rate $\alpha$ by a signal from infected cells, a proportion $p_1$ of them will develop into regulatory T cells, a proportion $p_2$ will become normal activated T cells able to destroy infected cells at rate $\mu_F$, and the remaining T cells will become autoreactive, in which case their threshold for activation by susceptible cells is reduced, and hence, they will be destroying both infected and susceptible host cells at rate $\mu_a$. The effect of regulatory T cells is in reducing the number of autoreactive T cells at rate $\delta$, and regulatory T cells are themselves assumed to be in a state of homeostasis. Finally, normal and autoreactive T cells produce IL-2 at rates $\sigma_1$ and $\sigma_2$, and IL-2 in turn facilitates proliferation of regular, normal and autoreactive T cells at rates $\rho_1$, $\rho_2$ and $\rho_3$, respectively. One should note that in light of experimental evidence suggesting the possibility of autoimmunity in the absence of B cells~\cite{Wolf96} and the fact that the development of antibodies can itself depend on prior T cell activation with bacteria~\cite{Wu10}, the above model does not take into account antibody response, but rather focuses on T cell dynamics.

As a first step in the analysis of stochastic effects in immune dynamics, we construct a CTMC model based on the ODE model (\ref{firstmodel}) using the methodology developed earlier in the context of modelling stochastic effects in epidemic and immunological models~\cite{allen1, brauer, Stirk2010a}. To this end, we introduce variables $X_1(t),\ldots,X_7(t)\in\{0,1,2,\ldots\}$ as discrete random variables representing the number of uninfected cells, infected cells, na\"ive T cells, regulatory T cells, normal activated T cells, autoreactive T cells, and interleukin-2 at time $t$, respectively. Let the initial condition be fixed as
\[
\textbf{X}_0=(X_1(0),\ldots,X_7(0))=(n_{10},n_{20},n_{30},n_{40},n_{50},n_{60},n_{70}).
\]
The probability of finding the system in the state $\textbf{n}=(n_1,n_2,n_3,n_4,n_5,n_6,n_7)$ with $n_i \in \{0, 1, 2, ...\}$ at time $t$ can be defined as
\[
P(\textbf{n},t)=\mbox{Prob}\{\textbf{X}(t)=\textbf{n}|\textbf{X}(0)=\textbf{X}_0\}.
\]
Let $\Delta t$ be sufficiently small such that $\Delta X_i(t)=X_i(t+\Delta t)-X_i(t) \in \{-1,0,1\}$ for $1 \leq i\leq 7$. The CTMC can then be formulated as a birth and death process in each of the variables~\cite{allen1}.
The infinitesimal transition probabilities corresponding to Figure~\ref{fig1} are as follows,
\begin{equation}\label{Prob}
	\mbox{Prob}(\Delta \textbf{X}=\bf i|\textbf{X}=\bf n)=
	\begin{cases}
	q_1\Delta t + o(\Delta t), \hspace{1cm} \textbf{i}=(1,0,0,0,0,0,0), \\
	q_2\Delta t + o(\Delta t), \hspace{1cm} \textbf{i}=(-1,0,0,0,0,0,0), \\
	q_3\Delta t + o(\Delta t), \hspace{1cm}  \textbf{i}=(-1,1,0,0,0,0,0), \\
	q_4\Delta t + o(\Delta t), \hspace{1cm} \textbf{i}=(0,0,1,0,0,0,0), \\
	q_5\Delta t + o(\Delta t), \hspace{1cm}  \textbf{i}=(0,0,-1,0,0,0,0), \\
	q_6\Delta t + o(\Delta t), \hspace{1cm}  \textbf{i}=(0,0,-1,0,1,0,0), \\
	q_7\Delta t + o(\Delta t), \hspace{1cm}  \textbf{i}=(0,0,-1,0,1,0,0), \\
	q_8\Delta t + o(\Delta t), \hspace{1cm}  \textbf{i}=(0,0,-1,0,0,1,0), \\
	q_9\Delta t + o(\Delta t), \hspace{1cm}  \textbf{i}=(0,-1,0,0,0,0,0), \\
	q_{10}\Delta t + o(\Delta t), \hspace{1cm}  \textbf{i}=(0,0,0,1,0,0,0), \\
	q_{11}\Delta t + o(\Delta t), \hspace{1cm}  \textbf{i}=(0,0,0,-1,0,0,0), \\
	q_{12}\Delta t + o(\Delta t), \hspace{1cm}  \textbf{i}=(0,0,0,0,1,0,0), \\
	q_{13}\Delta t + o(\Delta t), \hspace{1cm}  \textbf{i}=(0,0,0,0,-1,0,0), \\
	q_{14}\Delta t + o(\Delta t), \hspace{1cm}  \textbf{i}=(0,0,0,0,0,1,0), \\
	q_{15}\Delta t + o(\Delta t), \hspace{1cm}   \textbf{i}=(0,0,0,0,0,-1,0), \\
	q_{16}\Delta t + o(\Delta t), \hspace{1cm}   \textbf{i}=(0,0,0,0,0,0,1), \\
	q_{17}\Delta t + o(\Delta t), \hspace{1cm}   \textbf{i}=(0,0,0,0,0,0,-1), \\
	1-\sum\limits_{i=1}^{17} q_i\Delta t+ o(\Delta t), \hspace{1cm}  \textbf{i}=(0,0,0,0,0,0,0), \\
	o(\Delta t), \hspace{1cm}  \mbox{ otherwise, }
	\end{cases}
\end{equation}
where
\begin{align*}
	&q_1=b_1n_1+b_2n_1^2, \quad q_2=d_1n_1+d_2n_1^2+\mu_an_1n_6, \quad q_3=\beta n_1n_2, \quad q_4=\lambda_{in},\\
	&q_5=d_{in}n_3, \quad q_6=p_1\alpha n_2n_3, \quad q_7=p_2\alpha n_2n_3, \quad q_8=(1-p_1-p_2)\alpha n_2n_3,\\
	&q_9=(d_F+\mu_Fn_5+\mu_an_6)n_2, \quad q_{10}=\lambda_r+\rho_1n_4n_7, \quad q_{11}=d_rn_4, \quad q_{12}=\rho_2n_5n_7,\\
	&q_{13}=d_nn_5, \quad q_{14}=\rho_3n_6n_7, \quad q_{15}=(d_a+\delta n_4)n_6, \quad q_{16}=\sigma_1 n_5+\sigma_2 n_6, \quad q_{17}=d_in_7.
\end{align*}
Here, $b_1n_1+b_2n_1^2$ and $d_1n_1+d_2n_1^2$ are natural birth and death rates for uninfected cells with $b_1-d_1=r$ and $d_2-b_2=r/N$~\cite{allen1}.

The probabilities $P(\textbf{n},t)$ satisfy the following master equation (forward Kolmogorov equation)
\begin{align}\label{master equation}
	\dfrac{dP(\textbf{n},t)}{dt}=&\{(\varepsilon_1^--1)q_1+(\varepsilon_1^+-1)q_2+(\varepsilon_1^+\varepsilon_2^--1)q_3+(\varepsilon_3^--1)q_4+(\varepsilon_3^+-1)q_5 \nonumber \\
	&+(\varepsilon_3^+\varepsilon_4^--1)q_6+(\varepsilon_3^+\varepsilon_5^--1)q_7+(\varepsilon_3^+\varepsilon_6^--1)q_8+(\varepsilon_2^+-1)q_9+(\varepsilon_4^--1)q_{10} \nonumber \\
	&+(\varepsilon_4^+-1)q_{11}+(\varepsilon_5^--1)q_{12}+(\varepsilon_5^+-1)q_{13}+(\varepsilon_6^--1)q_{14}+(\varepsilon_6^+-1)q_{15} \nonumber \\
	&+(\varepsilon_7^--1)q_{16}+(\varepsilon_7^+-1)q_{17}\}P(\textbf{n},t).
\end{align}
where the operators $\varepsilon_i^\pm$ are defined as follows,
\[
	\varepsilon_i^\pm f(n_1,n_2,n_3,n_4,n_5,n_6,n_7,t)=f(n_1,...,n_i\pm 1,...,n_7,t),\mbox{ for each }1\leq i\leq 7, 
\]
and if $n_i<0$ for any $1\leq i\leq 7$, then $P(\textbf{n}, t)=0$.

\begin{table}
	\centering
	\caption{Possible state changes $\Delta\textbf{Y}$ during a small time interval $\Delta t$}
	\begin{tabular}{|l|l|l|}
		\hline
		i  & $(\Delta\textbf{Y})_i^T$ & Probability $P_i\Delta t$                 \\ \hline
		1  & $(1,0,0,0,0,0,0)$        & $(b_1Y_1+b_2{Y_1}^2)\Delta t$             \\ \hline
		2  & $(-1,0,0,0,0,0,0)$       & $(d_1Y_1+d_2{Y_1}^2+\mu_aY_6Y_1)\Delta t$ \\ \hline
		3  & $(-1,1,0,0,0,0,0)$       & $\beta Y_1Y_2\Delta t$                    \\ \hline
		4  & $(0,0,1,0,0,0,0)$        & $\lambda_{in}\Delta t$                    \\ \hline
		5  & $(0,0,-1,0,0,0,0)$       & $d_{in}Y_3\Delta t$                       \\ \hline
		6  & $(0,0,-1,1,0,0,0)$       & $p_1\alpha Y_3Y_2\Delta t$                \\ \hline
		7  & $(0,0,-1,0,1,0,0)$       & $p_2\alpha Y_3Y_2\Delta t$                \\ \hline
		8  & $(0,0,-1,0,0,1,0)$       & $(1-p_1-p_2)\alpha Y_3Y_2\Delta t$        \\ \hline
		9  & $(0,-1,0,0,0,0,0)$       & $(d_F+\mu_FY_5+\mu_aY_6)Y_2\Delta t$      \\ \hline
		10 & $(0,0,0,1,0,0,0)$        & $(\lambda_r+\rho_1Y_7Y_4)\Delta t$        \\ \hline
		11 & $(0,0,0,-1,0,0,0)$       & $d_rY_4\Delta t$                          \\ \hline
		12 & $(0,0,0,0,1,0,0)$        & $\rho_2Y_7Y_5\Delta t$                    \\ \hline
		13 & $(0,0,0,0,-1,0,0)$       & $d_nY_5\Delta t$                          \\ \hline
		14 & $(0,0,0,0,0,1,0)$        & $\rho_3Y_7Y_6\Delta t$                    \\ \hline
		15 & $(0,0,0,0,0,-1,0)$       & $(d_a+\delta Y_4)Y_6\Delta t$             \\ \hline
		16 & $(0,0,0,0,0,0,1)$        & $(\sigma_1 Y_5+\sigma_2 Y_6)\Delta t$     \\ \hline
		17 & $(0,0,0,0,0,0,-1)$       & $d_iY_7\Delta t$                          \\ \hline
		18 & $(0,0,0,0,0,0,0)$        & $1-\sum\limits_{i=1}^{17} P_i\Delta t$    \\ \hline
	\end{tabular}
    \label{possible changes}
%    \end{adjustwidth}
\end{table}

By solving this master equation, one can find the probability density function for this model. However, since this is a high-dimensional difference-differential equation, solving it is a very challenging task. Normally, the number of events occurring in a small time step in the CTMC model is extremely large, hence using the CTMC model for plotting stochastic trajectories is very computationally intensive~\cite{allen2}. A much more computationally efficient approach is to use chemical Langevin equations~\cite{gill1, gill2}, also known as It\^o stochastic differential equation (SDE) models, which provide very similar sample paths to those of the CTMC models~\cite{allen2}. While both It\^o and Stratonovich interpretations of stochastic calculus can be applied~\cite{oksen00}, in biological applications It\^o formulation is more frequently used due to its
non-anticipatory nature and a closer connection to numerical implementation~\cite{brau08,eallen,allen1}.

\subsection{Stochastic differential equation model}

To derive It\^o SDE model, let $\textbf{Y}(t)=(Y_1(t), Y_2(t), Y_3(t), Y_4(t), Y_5(t), Y_6(t), Y_7(t))$ be a continuous random vector for the sizes of various cell compartments at time $t$. Similar to the CTMC model, we assume that $\Delta t$ is small enough so that during this time interval at most one change can occur in state variables. These changes together with their probabilities are listed in Table~\ref{possible changes}, which is again based on Figure~\ref{fig1} and transitions in the CTMC model (\ref{Prob}). Using this table of possible state changes, one can compute the expectation vector and covariance matrix of $\Delta\textbf{Y}$ for sufficiently small $\Delta t$~\cite{allen2,allen4}. The expectation vector to order $\Delta t$ is given by
\[
\mathbb{E}(\Delta \textbf{Y})\approx\sum\limits_{i=1}^{17}P_i(\Delta\textbf{Y})_i\Delta t=\boldsymbol{\mu}\Delta t,
\]
where
\[
\boldsymbol{\mu}=
\begin{pmatrix}
P_1-P_2-P_3 \\
P_3-P_9 \\
P_4-P_5-P_6-P_7-P_8 \\
P_6+P_{10}-P_{11} \\
P_7+P_{12}-P_{13} \\
P_8+P_{14}-P_{15} \\
P_{16}-P_{17}
\end{pmatrix}
\]
is the drift vector, which can be easily seen to be identical to the right-hand side of the deterministic model (\ref{firstmodel}). The covariance matrix is obtained by keeping terms of order $\Delta t$ only, i.e.
\[
\begin{array}{l}
\mbox{cov}(\Delta\textbf{Y})=\mathbb{E}\left[(\Delta \textbf{Y})(\Delta \textbf{Y})^T\right]-\mathbb{E}\left[\Delta \textbf{Y}\right](\mathbb{E}\left[\Delta \textbf{Y}\right])^T\approx \mathbb{E}\left[(\Delta \textbf{Y})(\Delta \textbf{Y})^T\right]\\\\
=\sum\limits_{i=1}^{17}P_i(\Delta \textbf{Y})_i(\Delta \textbf{Y}_i)^T\Delta t=\Sigma \Delta t,
\end{array}
\]
where
\[
\resizebox{\linewidth}{!}{%
	$\displaystyle
	\Sigma=
	\begin{pmatrix}
	P_1+P_2+P_3 & -P_3 & 0 & 0 & 0 & 0 & 0 \\
	-P_3 & P_3+P_9 & 0 & 0 & 0 & 0 & 0 \\
	0 & 0 & P_4+P_5+P_6+P_7+P_8 & -P_6 & -P_7 & -P_8 & 0 \\
	0 & 0 & -P_6 & P_6+P_{10}+P_{11} & 0 & 0 & 0 \\
	0 & 0 & -P_7 & 0 & P_7+P_{12}+P_{13} & 0 & 0 \\
	0 & 0 & -P_8 & 0 & 0 & P_8+P_{14}+P_{15} & 0 \\
	0 & 0 & 0 & 0 & 0 & 0 & P_{16}+P_{17}
	\end{pmatrix}
	$}
\]
is a $7\times 7$ covariance matrix. To derive It\^o SDE model, we need to find a diffusion matrix $H$ defined according to $HH^T=\Sigma$. Although this matrix is not unique, different forms of this matrix give equivalent systems \cite{eallen,allen4}.

If one rewrites the covariance matrix $\Sigma$ in the form
\[
\Sigma=
\begin{pmatrix}
U & \bf{0} & \bf{0} \\
\bf{0} & W & \bf{0} \\
\bf{0} & \bf{0} & Z
\end{pmatrix},
\]
with
\[
U=
\begin{pmatrix}
P_1+P_2+P_3 & -P_3 \\
-P_3 & P_3+P_9 
\end{pmatrix},\hspace{0.5cm}Z=P_{16}+P_{17},
\]
and
\[
W=
\begin{pmatrix}
P_4+P_5+P_6+P_7+P_8 & -P_6 & -P_7 & -P_8 \\
-P_6 & P_6+P_{10}+P_{11} & 0 & 0 \\
-P_7 & 0 & P_7+P_{12}+P_{13} & 0 \\
-P_8 & 0 & 0 & P_8+P_{14}+P_{15}
\end{pmatrix},
\]
we can define three matrices $H_1$, $H_2$ and $H_3$ as follows,
\[
H_1=
\begin{pmatrix}
\sqrt{P_1+P_2} & -\sqrt{P_3} & 0 \\
0 & \sqrt{P_3} & \sqrt{P_9} 
\end{pmatrix},\hspace{0.5cm}H_3=
\sqrt{P_{16}+P_{17}},
\]
\[
H_2=
\begin{pmatrix}
\sqrt{P_4+P_5} & -\sqrt{P_6} & -\sqrt{P_7} & -\sqrt{P_8} & 0 & 0 & 0 \\
0 & \sqrt{P_6} & 0 & 0 & \sqrt{P_{10}+P_{11}} & 0 & 0 \\
0 & 0 & \sqrt{P_7} & 0 & 0 & \sqrt{P_{12}+P_{13}} & 0 \\
0 & 0 & 0 & \sqrt{P_8} & 0 & 0 & \sqrt{P_{14}+P_{15}}
\end{pmatrix}.
\]
Now if we consider
\[
H=
\begin{pmatrix}
H_1 & \bf{0} & \bf{0} \\
\bf{0} & H_2 & \bf{0} \\
\bf{0} & \bf{0} & H_3
\end{pmatrix},
\]
then $HH^T=\Sigma$, where $H$ is a $7\times 11$ matrix. The It\^o SDE model now has the form
\begin{equation}\label{SDE}
	\begin{cases}
		d\textbf{Y}(t)=\boldsymbol{\mu}dt+Hd\textbf{W}(t), \\ \textbf{Y}(0)=(A(0),F(0),T_{in}(0),T_{reg}(0),T_{nor}(0),T_{aut}(0),I(0))^T,
	\end{cases}
\end{equation}
\noindent
and $\textbf{W}(t)=[W_1(t),W_2(t),...,W_{11}(t)]^T$ is a vector of eleven independent Wiener processes~\cite{eallen}.

In order to make further analytical progress, we find an approximate probability density function for the model~(\ref{SDE}) as given by an approximate  solution of the master equation~\cite{eallen, kampen}. Let $P(\textbf{Y},t)$ be the probability density function of the model~(\ref{SDE}). Then $P(\textbf{Y},t)$ satisfies the following Fokker-Planck equation~\cite{eallen, gardiner} which is an approximation of the master equation
\[
	\begin{cases}
	\displaystyle{\dfrac{\partial P(\textbf{Y},t)}{\partial t}=-\sum\limits_{i=1}^{7}\dfrac{\partial}{\partial y_i}\left[\mu_iP(\textbf{Y},t)\right]+\dfrac{1}{2}\sum\limits_{i=1}^{7}\sum\limits_{j=1}^{7}\dfrac{\partial^2}{\partial y_i \partial y_j}\left[\Sigma_{ij}P(\textbf{Y},t)\right],}\\\\
	P(\textbf{Y},0)=\delta_7(\textbf{Y}-\textbf{Y}_0).\\
	\end{cases}
	\]
\noindent
By solving this PDE, one can find the probability density function of our model, but since this equation is high-dimensional and nonlinear, solving it analytically is impossible. Hence, we use another approach, a so-called system size expansion or van Kampen's $\Omega$-expansion~\cite{kampen}, which is a method for constructing a continuous approximation to a discrete stochastic model~\cite{Stirk2010a,Stirk2010b}, which allows one to study stochastic fluctuations around deterministic attractors~\cite{Black2009}.

\subsection{System size expansion}

In order to apply the van Kampen's approach, we consider fluctuations within a systematic expansion of the master equation for a large system size $\Omega$. Specifically, we write each $n_i(t)$ as a deterministic part of order $\Omega$ plus a fluctuation of order $\Omega^{1/2}$ as follows,
\begin{equation}\label{SSize}
n_i(t)=\Omega x_i(t)+\Omega^{1/2}\zeta_i(t),\hspace{0.5cm}i=1,\ldots,7,
\end{equation}
where $x_i(t)$ and $\zeta_i(t)$ are two continuous variables, and $\Omega x_i(t)=\mathbb{E}[n_i(t)]$. The probability density $P(\textbf{n},t)$ satisfying the master equation (\ref{master equation}) is now represented by the probability density $\Pi(\boldsymbol{\zeta},t)$, i.e. $\Pi(\boldsymbol{\zeta},t)=P(\textbf{n},t)=P\left(\Omega {\bf x}+\Omega^{1/2}\boldsymbol{\zeta},t\right)$, which implies
\begin{equation}\label{pi}
\displaystyle{\frac{dP(\textbf{n},t)}{dt}=\frac{\partial \Pi}{\partial t}-\sum\limits_{i=1}^{7}\Omega^{1/2}\dfrac{dx_i}{dt}\frac{\partial \Pi}{\partial \zeta_i}.}
\end{equation}
To expand the master equation (\ref{master equation}) in a power series in $\Omega^{-1/2}$, we use the following expansion for the step operators
\begin{equation}\label{expansion}
\displaystyle{\varepsilon_i^{\pm}=1\pm \Omega^{-1/2}\dfrac{\partial}{\partial \zeta_i}+\frac{1}{2}\Omega^{-1}\frac{\partial^2}{\partial \zeta_i^2}\pm \cdots.}
\end{equation}
Substituting expressions~(\ref{pi}) and (\ref{expansion}) into the master equation (see {\bf Supplementary Material} for details) and collecting terms of order $\Omega^{1/2}$ yields the following deterministic model for macroscopic behaviour
\begin{align}\label{secondmodel}
	\begin{split}
		\frac{dx_1}{dt}&=b_1x_1+\widetilde{b}_2x_1^2-d_1x_1-\widetilde{d}_2x_1^2-\widetilde{\beta}x_1x_2-\widetilde{\mu}_a x_1x_6,\\
		\frac{dx_2}{dt}&=\widetilde{\beta}x_1x_2-d_Fx_2-\widetilde{\mu}_Fx_2x_5-\widetilde{\mu}_a x_2x_6,\\
		\frac{dx_3}{dt}&=\widetilde{\lambda}_{in}-d_{in}x_3-\widetilde{\alpha}x_2x_3,\\
		\frac{dx_4}{dt}&=\widetilde{\lambda}_r-d_rx_4+p_1\widetilde{\alpha}x_2x_3+\widetilde{\rho}_1x_4x_7,\\
		\frac{dx_5}{dt}&=p_2\widetilde{\alpha}x_2x_3-d_nx_5+\widetilde{\rho}_2 x_5x_7,\\
		\frac{dx_6}{dt}&=(1-p_1-p_2)\widetilde{\alpha}x_2x_3-d_ax_6-\widetilde{\delta}x_4x_6+\widetilde{\rho}_3x_6x_7,\\
		\frac{dx_7}{dt}&=\sigma_1x_5+\sigma_2x_6-d_ix_7,
	\end{split}
\end{align}
where
\begin{align*}
	&b_2=\dfrac{\widetilde{b}_2}{\Omega}, \quad d_2=\dfrac{\widetilde{d}_2}{\Omega}, \quad \beta=\dfrac{\widetilde{\beta}}{\Omega}, \quad \mu_a=\dfrac{\widetilde{\mu}_a}{\Omega}, \quad \mu_F=\dfrac{\widetilde{\mu}_F}{\Omega}, \quad \alpha=\dfrac{\widetilde{\alpha}}{\Omega}, \quad \delta=\dfrac{\widetilde{\delta}}{\Omega},\\
	&\rho_i=\dfrac{\widetilde{\rho}_i}{\Omega}, \quad i=1,2,3, \quad \lambda_{in}=\widetilde{\lambda}_{in}\Omega, \quad \lambda_{r}=\widetilde{\lambda}_{r}\Omega.
\end{align*}

Model~(\ref{secondmodel}) has been analysed in (Fatehi et al. 2017), and it can have at most four biologically feasible steady states. The first one, a disease-free steady state, is given by
\[
S^{\ast}_1=\left(\dfrac{b_1-d_1}{\widetilde{d}_2-\widetilde{b}_2},0,\dfrac{\widetilde{\lambda}_{in}}{d_{in}},\dfrac{\widetilde{\lambda}_{r}}{d_r},0,0,0\right),
\]
and it is stable if $d_F>\widetilde{\beta}$. The second and third steady states can be found as
\[
S^{\ast}_2=\left(0,0,\dfrac{\widetilde{\lambda}_{in}}{d_{in}},x^{\ast}_4,0,\dfrac{d_i\left(d_a+\widetilde{\delta}x^{\ast}_4\right)}{\widetilde{\rho}_3\sigma_2},\dfrac{d_a+\widetilde{\delta}x^{\ast}_4}{\widetilde{\rho}_3}\right),
\]
and
\[
S^{\ast}_3=\left(\dfrac{\widetilde{\rho}_3\sigma_2(b_1-d_1)-\widetilde{\mu}_ad_i\left(d_a+\widetilde{\delta}x^{\ast}_4\right)}{\widetilde{\rho}_3\sigma_2\left(\widetilde{d}_2-\widetilde{b}_2\right)},0,\dfrac{\widetilde{\lambda}_{in}}{d_{in}},x^{\ast}_4,0,\dfrac{d_i\left(d_a+\widetilde{\delta}x^{\ast}_4\right)}{\widetilde{\rho}_3\sigma_2},\dfrac{d_a+\widetilde{\delta}x^{\ast}_4}{\widetilde{\rho}_3}\right),
\]
where $x^{\ast}_4$ satisfies the following quadratic equation
\begin{equation}\label{treg equation}
	\widetilde{\rho}_1\widetilde{\delta}\left(x^{\ast}_4\right)^2+\left(\widetilde{\rho}_1d_a-\widetilde{\rho}_3d_r\right)x^{\ast}_4+\widetilde{\rho}_3\widetilde{\lambda}_r=0.
\end{equation}
These steady states are stable, provided
\[
\begin{array}{l}
\dfrac{\sigma_2}{\widetilde{\mu}_ad_i}K<\dfrac{d_a+\widetilde{\delta}x^{\ast}_4}{\widetilde{\rho}_3}< \dfrac{d_n}{\widetilde{\rho}_2},\quad \widetilde{\delta} \widetilde{\rho}_1(x^{\ast}_4)^2>\widetilde{\lambda}_r\widetilde{\rho}_3,\\\\
\widetilde{\rho}_3\widetilde{\lambda}_r^2+\widetilde{\rho}_3d_i\widetilde{\lambda}_rx^{\ast}_4-\widetilde{\rho}_3d_id_a(x^{\ast}_4)^2-\widetilde{\delta}(\widetilde{\rho}_1d_a+\widetilde{\rho}_3d_i)(x^{\ast}_4)^3-\widetilde{\rho}_1\widetilde{\delta}^2(x^{\ast}_4)^4>0,
\end{array}
\]
where $K=1$ for $S^{\ast}_2$, and $K=\left(\widetilde{\beta}-d_F\right)/\left(1+\widetilde{\beta}\right)$ for $S^{\ast}_3$. Biologically, the steady state $S^{\ast}_2$ represents the death of organ cells, while $S^{\ast}_3$ corresponds to an autoimmune regime.

The last steady state $S^{\ast}_4$ has all of its components positive and corresponds to the state of chronic infection.

At the next order, stochastic fluctuations are determined by linear stochastic processes, hence, this is known as a linear noise approximation~\cite{kampen, Wallace2013}. The dynamics of these fluctuations is described by the following linear Fokker-Planck equation
\begin{equation}\label{linearFPE}
	\dfrac{\partial\Pi(\boldsymbol{\zeta},t)}{\partial t}=-\sum\limits_{i,j}A_{ij}\dfrac{\partial}{\partial \zeta_i}(\zeta_j\Pi)+\dfrac{1}{2}\sum\limits_{i,j}B_{ij}\dfrac{\partial^2\Pi}{\partial\zeta_i \partial\zeta_j},
\end{equation}
where $A$ is the Jacobian matrix of system (\ref{secondmodel})
\[
\resizebox{\linewidth}{!}{%
	$\displaystyle
	A=
	\begin{pmatrix}
	b_1+2\widetilde{b}_2x_1-d_1-2\widetilde{d}_2x_1-\widetilde{\mu}_ax_6-\widetilde{\beta}x_2 & -\widetilde{\beta}x_1 & 0 & 0 & 0 & -\widetilde{\mu}_ax_1 & 0 \\
	\widetilde{\beta}x_2 & \widetilde{\beta}x_1-d_F-\widetilde{\mu}_Fx_5-\widetilde{\mu}_ax_6 & 0 & 0 & -\widetilde{\mu}_Fx_2 & -\widetilde{\mu}_ax_2 & 0 \\
	0 & -\widetilde{\alpha}x_3 & -d_{in}-\widetilde{\alpha}x_2 & 0 & 0 & 0 & 0 \\
	0 & p_1\widetilde{\alpha}x_3 & p_1\widetilde{\alpha}x_2 & \widetilde{\rho}_1x_7-d_r & 0 & 0 & \widetilde{\rho}_1x_4\\
	0 & p_2\widetilde{\alpha}x_3 & p_2\widetilde{\alpha}x_2 & 0 & \widetilde{\rho}_2x_7-d_n & 0 & \widetilde{\rho}_2x_5\\
	0 & (1-p_1-p_2)\widetilde{\alpha}x_3 & (1-p_1-p_2)\widetilde{\alpha}x_2 & -\widetilde{\delta}x_6 & 0 & \widetilde{\rho}_3x_7-d_a-\widetilde{\delta}x_4 & \widetilde{\rho}_3x_6\\
	0 & 0 & 0 & 0 & \sigma_1 & \sigma_2 & -d_i 
	\end{pmatrix},
	$}
\]
and $B$ is a $7\times 7$ symmetric matrix given by
\begin{center}
	$B_{ij}$ = $\begin{cases} b_1x_1+\widetilde{b}_2x_1^2+d_1x_1+\widetilde{d}_2x_1^2+\widetilde{\beta}x_1x_2+\widetilde{\mu}_a x_1x_6, &\mbox{ if } (i,j) = (1,1),\\ 
	\widetilde{\beta}x_1x_2+d_Fx_2+\widetilde{\mu}_Fx_2x_5+\widetilde{\mu}_a x_2x_6, & \mbox{ if } (i,j) = (2,2),\\
	\widetilde{\lambda}_{in}+d_{in}x_3+\widetilde{\alpha}x_2x_3, & \mbox{ if } (i,j) = (3,3),\\
	\widetilde{\lambda}_r+d_rx_4+p_1\widetilde{\alpha}x_2x_3+\widetilde{\rho}_1x_4x_7, & \mbox{ if } (i,j) = (4,4),\\
	p_2\widetilde{\alpha}x_2x_3+d_nx_5+\widetilde{\rho}_2 x_5x_7, & \mbox{ if } (i,j) = (5,5),\\
	(1-p_1-p_2)\widetilde{\alpha}x_2x_3+d_ax_6+\widetilde{\delta}x_4x_6+\widetilde{\rho}_3x_6x_7, & \mbox{ if } (i,j) = (6,6),\\
	\sigma_1x_5+\sigma_2x_6+d_ix_7, & \mbox{ if } (i,j) = (7,7),\\
	-\widetilde{\beta}x_1x_2, & \mbox{ if } (i,j) = (1,2)\mbox{ or }(2,1),\\
	-p_1\widetilde{\alpha}x_2x_3, & \mbox{ if } (i,j) = (3,4)\mbox{ or }(4,3),\\
	-p_2\widetilde{\alpha}x_2x_3, & \mbox{ if } (i,j) = (3,5)\mbox{ or }(5,3),\\
	-(1-p_1-p_2)\widetilde{\alpha}x_2x_3, & \mbox{ if } (i,j) = (3,6)\mbox{ or }(6,3),\\
	0, & \mbox{ otherwise.}
	\end{cases}$
\end{center}
Since the Fokker-Planck equation~(\ref{linearFPE}) is linear, the probability density $\Pi(\boldsymbol{\zeta},t)$ is Gaussian, and hence, just the first two moments are enough to characterise it~\cite{hayot,Pahle2012}. Due to the way the system size expansion was introduced in (\ref{SSize}), the mean values of fluctuations for all variables are zero, i.e. $\langle \zeta_i(t)\rangle=0$ for all $1\leq i\leq 7$, while the covariance matrix $\Xi$ with $\Xi_{ij}=\langle\zeta_i(t)\zeta_j(t)\rangle-\langle\zeta_i(t)\rangle\langle\zeta_j(t)\rangle=\langle\zeta_i(t)\zeta_j(t)\rangle$ satisfies the following equation~\cite{kampen, Pahle2012}
\begin{equation}\label{covariance equation}
	\partial_t\Xi=A\Xi+\Xi A^T+B,
\end{equation}
where $A^T$ is the transpose of $A$.

We are mainly interested in the dynamics of fluctuations when the oscillations of the deterministic model have died out, and the system is in a stationary state, i.e. the fluctuations take place around the steady states~\cite{Black2009}. If the model~(\ref{secondmodel}) tends to a steady state as $t\rightarrow\infty$, then in the equation~(\ref{linearFPE}) one can substitute the values of $x_i$'s with the corresponding constant components of that steady state to study the fluctuations around it, as described by the linear Fokker-Planck equation. At any steady state, the covariance matrix $\Xi$ is independent of time, and the fluctuations are described by a Gaussian distribution with the zero mean and the stationary covariance satisfying the equation
\[
A\Xi+\Xi A^T+B=0.
\]
\begin{table}
	\centering
	\caption{Table of parameters}
	\begin{tabular}{|l|l|l|l|}
		\hline
		Parameter                  & value           & Parameter            & value             \\ \hline
		$b_1$                      & 2.5             & $d_r$                & 0.8               \\
		\hline 
		$\widetilde{b}_2$          & 0.1             & $p_1$                & 0.4               \\ 
		\hline
		$d_1$                      & 0.5             & $\widetilde{\rho}_1$ & $10/9$   \\ 
		\hline
		$\widetilde{d}_2$          & 0.2             & $p_2$                & 0.4               \\ 
		\hline
		$\widetilde{\beta}$        & 0.1             & $d_n$                & 2                 \\
		\hline 
		$\widetilde{\mu}_a$        & $40/9$ & $\widetilde{\rho}_2$ & $4/45$   \\
		\hline 
		$d_F$                      & 2.2             & $d_a$                & 0.002             \\
		\hline
		$\widetilde{\mu}_F$        & $4/3$  & $\widetilde{\delta}$ & $1/4500$ \\
		\hline
		$\widetilde{\lambda}_{in}$ & 18              & $\widetilde{\rho}_3$ & $2/9$    \\
		\hline 
		$d_{in}$                   & 2               & $\sigma_1$           & 0.3               \\
		\hline 
		$\widetilde{\alpha}$       & 0.04            & $\sigma_2$           & 0.4               \\
		\hline 
		$\widetilde{\lambda}_r$    & 108             & $d_i$                & 1.2               \\ \hline
	\end{tabular}
\label{parameter table}
\end{table}
\noindent In order to be able to relate the results of this analysis to simulations, it is convenient to express the covariance matrix in terms of actual numbers of cells in each compartment, rather than deviations from stationary values. To this end, we instead use the covariance matrix $C$ defined as
$C_{ij}=\langle(n_i-\langle n_i\rangle)(n_j-\langle n_j\rangle)\rangle$, which, in light of the relation $C_{ij}=\Omega\Xi_{ij}$, satisfies the following Lyapunov equation~\cite{Pahle2012}
\begin{equation}\label{lyapunov}
	AC+CA^T+\Omega B=0.
\end{equation}
This equation can be solved numerically for each of the stable steady states to determine the variance of fluctuations around that steady state depending on system parameters.

\section{Results}

To simulate the dynamics of the model, we solve the system (\ref{SDE}) numerically using the Euler-Maruyama method with parameter values given in Table~\ref{parameter table}, and $\Omega=1000$. The initial condition is chosen to be of the form
\begin{equation}\label{initial condition}
(x_1(0),x_2(0),x_3(0),x_4(0),x_5(0),x_6(0),x_7(0))=(18,2,7.2,6.3,0,0,0),
\end{equation}
which corresponds to a small number of host cells being initially infected.  

\begin{figure}[h]
    \begin{center}
    \includegraphics[width=12cm]{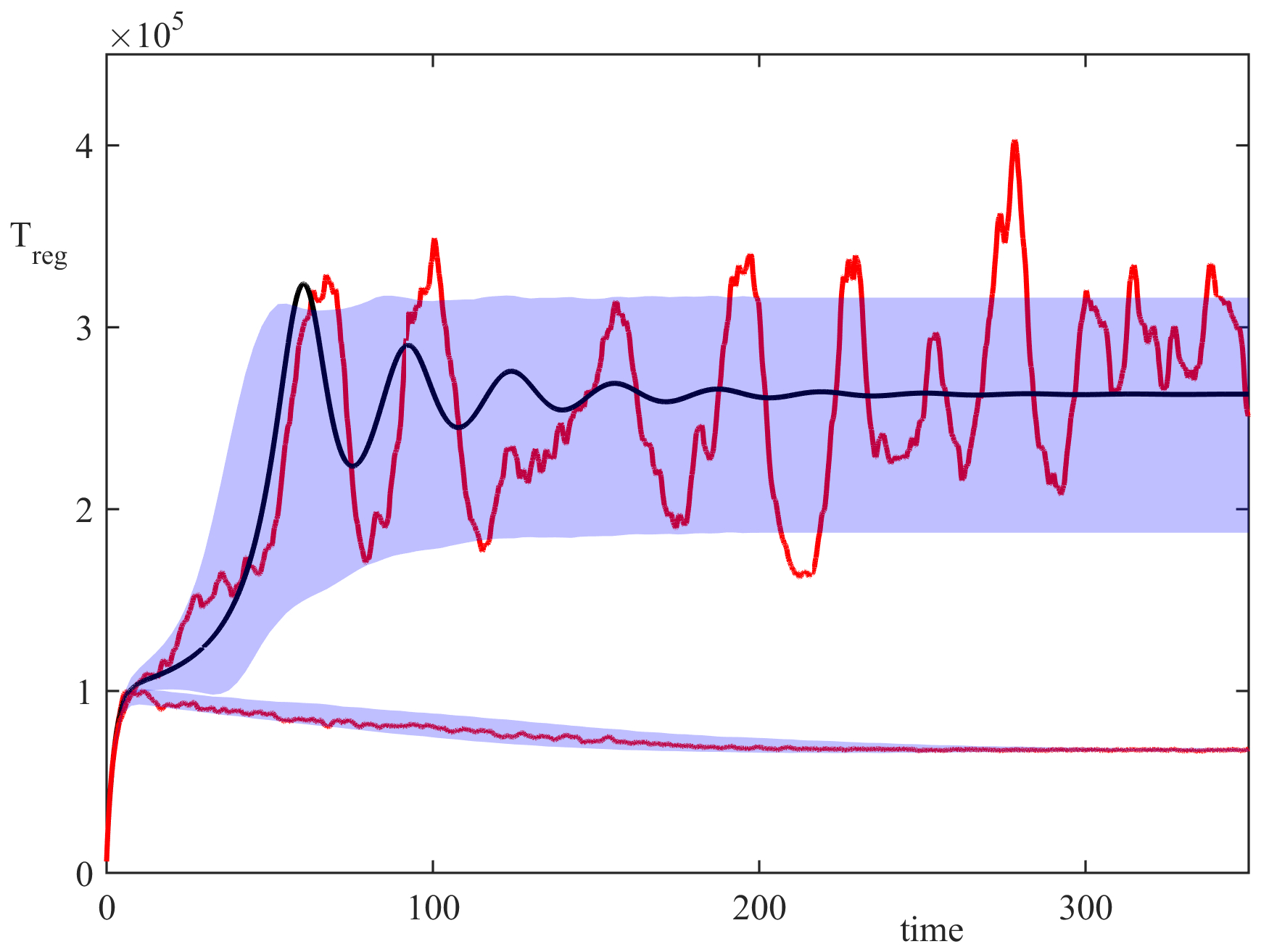}% This is a *.jpg file
    \end{center}
    \caption{Numerical simulation of the model (\ref{SDE}) with parameter values from Table~\ref{parameter table}, $\sigma_2=1$, and the initial condition~(\ref{initial condition}). Red curves are two sample paths that have entered the basins of attraction of $S^{\ast}_1$ or $S^{\ast}_3$, black curve is the deterministic trajectory from (\ref{firstmodel}), and the shaded areas indicate the regions of one standard deviation from the mean.}
    \label{fig2}
    \end{figure}

Figure~\ref{fig2} shows the results of 20000 simulations with the initial condition~(\ref{initial condition}) and $\sigma_2=1$. In the deterministic model~(\ref{secondmodel}), for $\sigma_2=1$ both steady states $S^{\ast}_1$ (disease-free) and $S^{\ast}_3$ (autoimmune state) are stable, but with the initial condition~(\ref{initial condition}) the system is in the basin of attraction of $S^{\ast}_3$. In the stochastic model, the majority of trajectories also enter the attraction region of $S^{\ast}_3$, but a small proportion of them went into the basin of attraction of $S^{\ast}_1$. This figure illustrates a single stochastic path around $S^{\ast}_1$, and a single stochastic path around $S^{\ast}_3$, together with the deterministic trajectory. These individual solutions indicate that whilst deterministically, the system exhibits decaying oscillations around $S^{\ast}_3$, the same behaviour is observed in the stochastic simulations only upon taking an average of a very
\newpage
\noindent large number of simulations. At the same time, individual realisations exhibit sustained stochastic oscillations in a manner similar to that observed in models of stochastic amplification in epidemics~\cite{Alonso2007,Kuske2007}. Figure~\ref{fig2} also illustrates the size of areas of one standard deviation from the mean for trajectories in the basins of attraction $S^{\ast}_1$ and $S^{\ast}_3$, in which individual stochastic trajectories may exhibit stochastic oscillations~\cite{Conway2011, Reyn12}.

\begin{figure}
    \begin{center}
    \includegraphics[width=14cm]{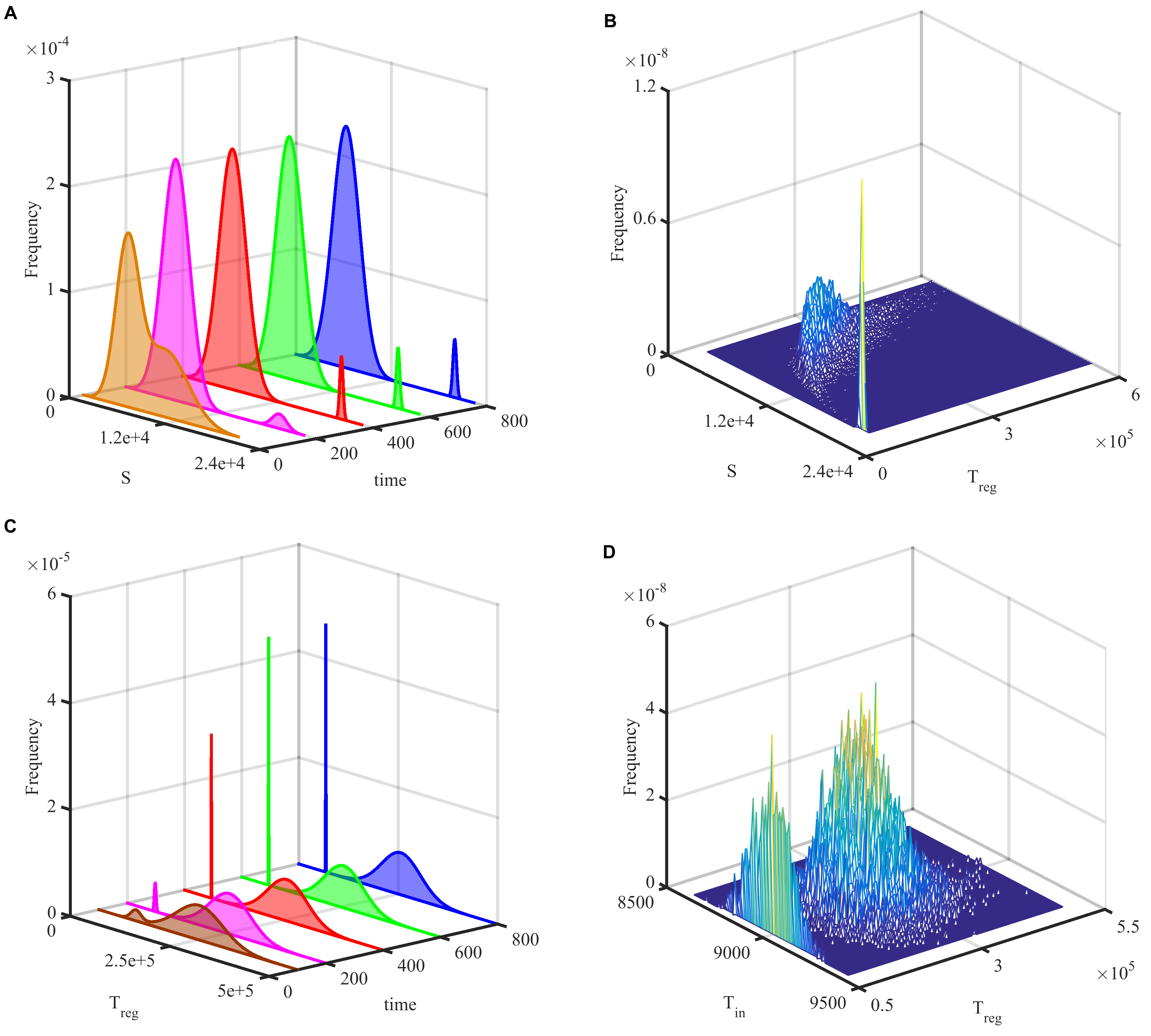}% This is a *.jpg file
    \end{center}
	\caption{Probability distribution of solutions out of 20000 simulations. (A) and (B) with parameters from Table~\ref{parameter table}, but $\sigma_2=1$ and the initial condition (\ref{initial condition}). (C) and (D) with parameters from Table~\ref{parameter table} and the initial condition (\ref{initial condition 1}). In (A) and (C), the probability histogram is fit to a bimodal normal distribution at different times. (B) and (D) illustrate stationary joint probability histograms.}
	\label{fig3}
\end{figure}

Figures~\ref{fig3}A and B show temporal evolution of the probability distribution in the case of bi-stability between the steady states $S^{\ast}_1$ and $S^{\ast}_3$, as illustrated in Figure~\ref{fig2}. They indicate that after some initial transient, the system reaches a stationary bimodal normal distribution. The width of the probability distribution around each stable steady state, as described by its variance or standard deviation, gives the size of fluctuations around this steady state observed in individual stochastic realisations, as is shown in Fig.~\ref{fig2}. Similar behaviour has been observed in stochastic realisations of other deterministic models with bi-stability~\cite{bruna14,earn13,hufton16}. For the parameter values given in Table~\ref{parameter table}, the deterministic system exhibits a bi-stability between $S^{\ast}_1$ and $S^{\ast}_2$, and with the initial condition
\begin{equation}\label{initial condition 1}
(x_1(0),x_2(0),x_3(0),x_4(0),x_5(0),x_6(0),x_7(0))=(18,9,7.2,6.3,0,0,0),
\end{equation}
it is in the basin of attraction of $S^{\ast}_2$. Due to stochasticity, the stationary probability distribution in this case is also bimodal, with the majority of solutions being distributed around $S^{\ast}_2$, and a very small number being centred around $S^{\ast}_1$, as can be seen in Figures~\ref{fig3}C and D. Increasing the system size $\Omega$ is known to result in the bimodal distribution becoming unimodal due to the size of fluctuations scaling as $\Omega^{-1/2}$, which results in a reduced variability in trajectories~\cite{Black2012,hufton16}, and the same conclusion holds for the system (\ref{SDE}).

\begin{figure}
    \begin{center}
    \includegraphics[width=14cm]{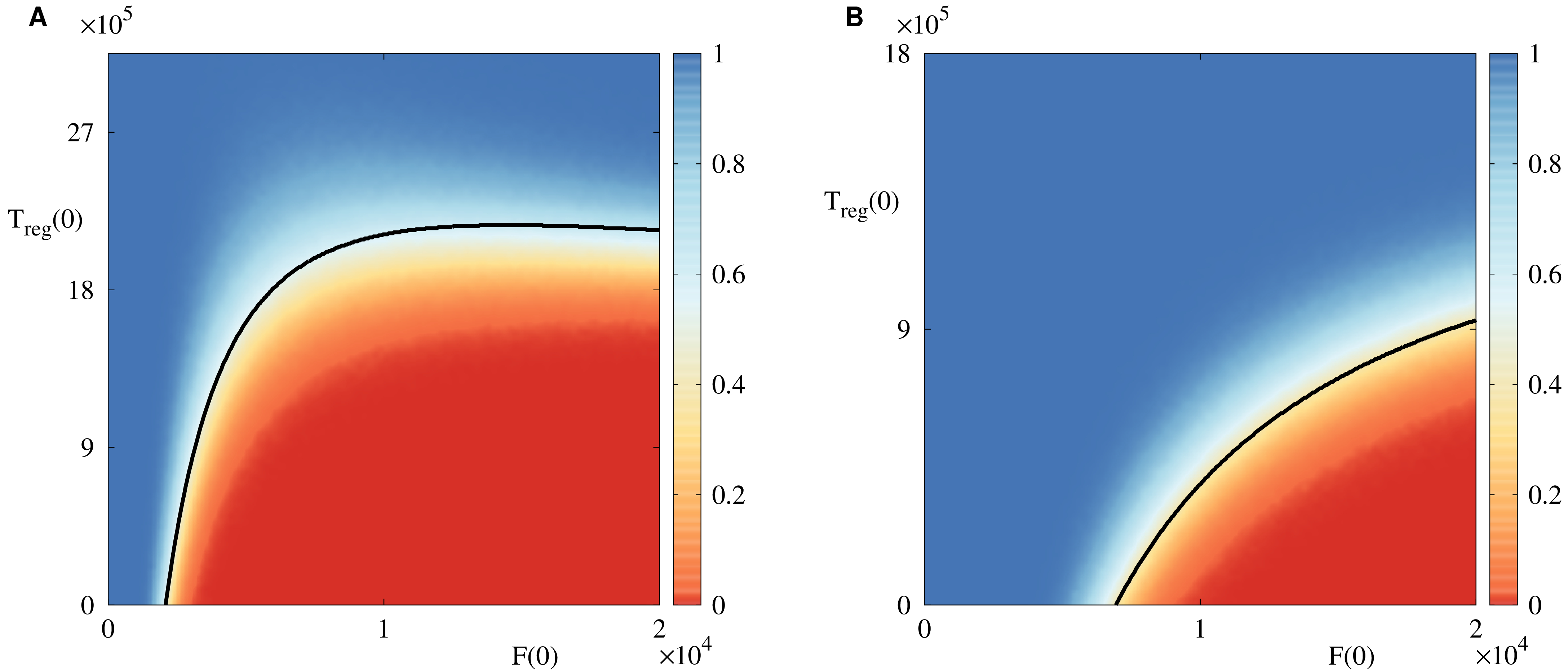}% This is a *.jpg file
    \end{center}
	\caption{Probability of solution entering and staying in the basin of attraction of the disease-free steady state $S^{\ast}_1$ in the bi-stability regime with $A(0)=18000$ and $T_{in}(0)=7200$. Black curves are the boundaries between different basins of attraction in the deterministic model. (A) With parameter values from Table~\ref{parameter table}, $\widetilde{\lambda}_r=45$ and $\widetilde{\mu}_a=10/9$, in the region below the black curve, the deterministic model exhibits a periodic solution around $S^{\ast}_3$, and above this curve is the deterministic basin of attraction of $S^{\ast}_1$. (B) With parameter values from Table~\ref{parameter table}, area below the black curve is the basin of attraction of $S^{\ast}_2$, and above it is again the basin of attraction of $S^{\ast}_1$.}
	\label{fig4}
\end{figure}

To gain better insights into the role of initial conditions, in Figure~\ref{fig4} we fix all parameter values, and vary initial numbers of infected cells and regulatory T cells.
For the parameter combination illustrated in Figure~\ref{fig4}A, the deterministic model exhibits a bi-stability between a stable disease-free steady state $S^{\ast}_1$ and a periodic oscillation around the state $S^{\ast}_3$, which biologically corresponds to an autoimmune regime. In the deterministic case, the black boundary provides a clear separation of the basins of attraction of these two dynamical states, in a manner similar to that investigated recently in the context of within-cell dynamics of RNA interference~\cite{Neofytou2017}. For stochastic simulations, the colour indicates the probability of the solution going to a disease-free state $S^{\ast}_1$, and it shows that even in the case where deterministically the system is in the basin of attraction of one of the states, there is a non-zero probability that it will actually end up at another state, with this probability varying smoothly across the deterministic basin boundary. This figure suggests that if the initial number of infected cells is sufficiently small, or if the number of regulatory T cells is sufficiently large, the system tends to clear the infection and approach the disease-free state. On the contrary, for higher numbers of infected cells and lower numbers of regulatory cells, autoimmune regime appears to be a more likely outcome. Qualitatively similar behaviour is observed for another combination of parameters illustrated in Figure~\ref{fig4}B, in which case the deterministic system has a bi-stability between a disease-free steady state $S^{\ast}_1$, and a state $S^{\ast}_2$ which represents the death of host cells.

\begin{figure}
    \begin{center}
    \includegraphics[width=14cm]{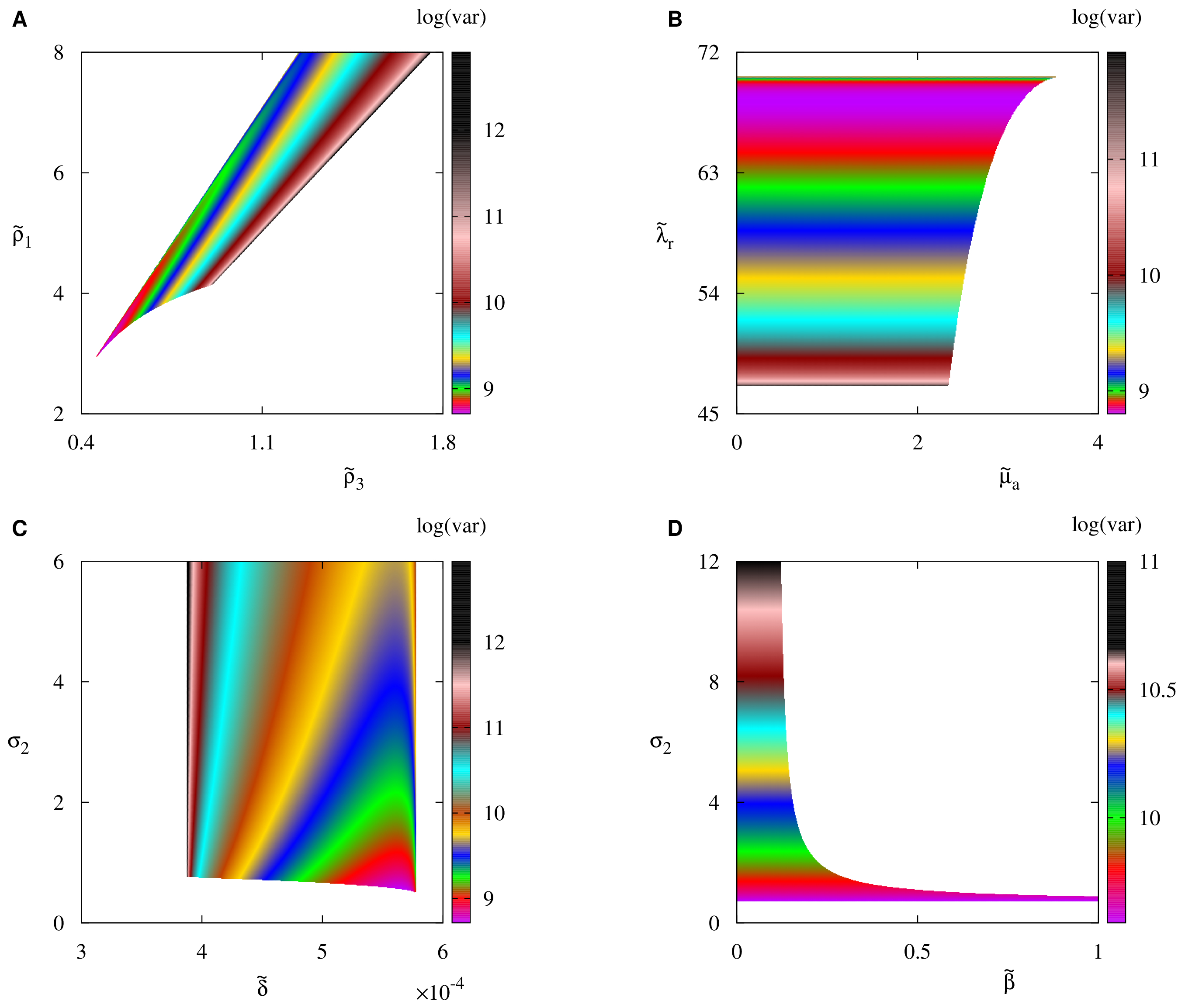}% This is a *.jpg file
    \end{center}
	\caption{Variance of the number of regulatory T cells $T_{reg}$ with parameter values from Table~\ref{parameter table}. Coloured regions indicate areas in respective parameter planes in which the autoimmune steady state $S^{\ast}_3$ is deterministically stable.}
	\label{fig5}
\end{figure}

In order to understand how biological parameters affect the size of fluctuations around steady states, in Figure~\ref{fig5} we explore several parameter planes by first identifying parameter regions where the deterministic system has a stable steady state $S^{\ast}_3$, and then for each combination of parameters inside these regions, we use the Bartels-Stewart method~\cite{BS72,Ham82}  to numerically solve the Lyapunov equation~(\ref{lyapunov}) and compute the variance in the number of regulatory T cell when the deterministic model is at the steady state $S^{\ast}_3$. The value of variance gives the square of the magnitude of oscillations observed in individual stochastic realisations. One should note that getting closer to the deterministic boundary of stability of $S^{\ast}_3$ increases the stochastic variance of fluctuations around this steady state. The reason for this is that closer parameters are to the deterministic stability boundary, the less stable is the steady state, hence the larger is the amplitude of stochastic oscillations around it. Moreover, the variance increases with the rate of production of IL-2 by autoreactive T cells and the rate at which regulatory T cells suppress autoreactive T cells; it decreases with the higher rate of production of regulatory T cells, and it appears to not depend on the rate at which autoreactive T cells destroy infected cells, or on the infection rate.

\section{Discussion}

In this paper we have analysed stochastic aspects of immune response against a viral infection with account for the populations of T cells with different activation thresholds, as well as cytokines mediating T cell activity. The CTMC model has provided an exact master equation, for which we applied a van Kampen's expansions to derive a linear Fokker-Planck equation that characterises fluctuations around the deterministic solutions. We have also explored actual stochastic trajectories of the system by deriving an SDE model and solving it numerically.

One biologically important aspect we have looked at is the influence of stochasticity on the dynamics of the system in the case where deterministically it exhibits a bi-stability between either two steady states, or a steady state and a periodic solution. In such a situation, bi-stability in the deterministic version of the model translates in the stochastic case into a stationary bimodal distribution for the probability density. To obtain further insights into details of how stochasticity affects bi-stability, we have investigated how for the fixed parameter values time evolution of the system changes depending on the initial numbers of the regulatory T cells and infected cells.

Our analysis reinforces the need to distinguish mean dynamics from individuals realisations: where in the deterministic case the system can approach a stable steady state (which represents mean behaviour of a very large number of simulations), individual realisations can exhibit sustained stochastic oscillations around that steady state, as we have seen in numerical simulations. Since in the clinical or laboratory setting one is usually dealing with single measurements of some specific biological quantities rather than their averaged values, the stochastic oscillations exhibited by our model may quite well explain observed variability in the measured levels of infection or T cell populations. To better understand the magnitude of stochastic fluctuations around the deterministic steady states, we have solved the Lyapunov equation, which has provided us with a quantitative information on the dependence of variance of fluctuations on system parameters.

There are several directions in which the work presented in this paper can be extended. In terms of fundamental immunology, the model can be made more realistic by including additional effects, such as the control of IL-2 secretion by regulatory T cells~\cite{Burr06}, or the memory T cells~\cite{Skap05,Antia05}. Whilst we have used numerical simulations to compute the probability of attraction to a given steady state in the case of bi-stability, one could approach the same problem theoretically from the perspective of computing extinction probability within the framework of the CTMC model~\cite{allen2, allen3}. The van Kampen's system size expansion could yield an expression for the power spectrum, which allows one to compute the peak frequency and amplification~\cite{Black2009, Black2010, Mckane2005, Alonso2007}. From a practical perspective, future work could focus on validating theoretical results presented in this paper using experimental measurements of the progress of autoimmune disease in animal hosts, with experimental autoimmune uveoretinitis (EAU), an autoimmune inflammation in the eyes, being one interesting possibility. In one such recent experiment, all animals were genetically identical C57BL/6 mice, but once the EAU was induced in them through inoculation, the autoimmune disease then progressed at slightly different rates ~\cite{Bold1}, (Boldison and Nicholson, 2012) and the measured variability in the numbers of infected cells and T cell responses could be compared to theoretical estimates of the variance as predicted by our model. From a clinical perspective, comparison of variance in the measured populations of different cells with the model conclusions will facilitate an efficient parameter identification and provide a set of prognostic criteria for the progress of autoimmunity, which can be used for risk stratification and assessment of patients with autoimmune disease.

\section*{Conflict of Interest Statement}
The authors declare that the research was conducted in the absence of any commercial or financial relationships that could be construed as a potential conflict of interest.

\section*{Author Contributions}

YK and KB designed the model; FF, SK and AR performed the analysis and simulations, and produced the figures. All authors drafted and edited the manuscript.

\section*{Funding}

FF has been sponsored by the Chancellor's Award from Sussex University. AR acknowledges funding for her PhD studies from the Engineering and
Physical Sciences Research Council (EPSRC) through iCASE Award EP/N509784/1.

\section*{Supplementary Material}
 The Supplementary Material for this article is attached. 

\bibliographystyle{ieeetr}
\bibliography{farzad}

\newpage

\section{Supplementary Material}
\subsection*{\LARGE Derivation of the van Kampen's system size expansion}
As described in Section 2.2 of the main text, the CTMC model based on the transitions probabilities yields the following master equation:
\begin{align}
\dfrac{dP(\textbf{n},t)}{dt}=&\{(\varepsilon_1^--1)q_1+(\varepsilon_1^+-1)q_2+(\varepsilon_1^+\varepsilon_2^--1)q_3+(\varepsilon_3^--1)q_4+(\varepsilon_3^+-1)q_5 \nonumber \\
&+(\varepsilon_3^+\varepsilon_4^--1)q_6+(\varepsilon_3^+\varepsilon_5^--1)q_7+(\varepsilon_3^+\varepsilon_6^--1)q_8+(\varepsilon_2^+-1)q_9+(\varepsilon_4^--1)q_{10} \nonumber \\
&+(\varepsilon_4^+-1)q_{11}+(\varepsilon_5^--1)q_{12}+(\varepsilon_5^+-1)q_{13}+(\varepsilon_6^--1)q_{14}+(\varepsilon_6^+-1)q_{15} \nonumber \\
&+(\varepsilon_7^--1)q_{16}+(\varepsilon_7^+-1)q_{17}\}P(\textbf{n},t),
\end{align}
where $\textbf{n}=(n_1,n_2,n_3,n_4,n_5,n_6,n_7)$ is the current state of the system, the coefficients $q_i$ are given by
\begin{align*}
	&q_1=b_1n_1+b_2n_1^2, \quad q_2=d_1n_1+d_2n_1^2+\mu_an_6n_1, \quad q_3=\beta n_1n_2, \quad q_4=\lambda_{in},\\
	&q_5=d_{in}n_3, \quad q_6=p_1\alpha n_3n_2, \quad q_7=p_2\alpha n_3n_2, \quad q_8=(1-p_1-p_2)\alpha n_3n_2,\\
	&q_9=(d_F+\mu_Fn_5+\mu_an_6)n_2, \quad q_{10}=\lambda_r+\rho_1n_7n_4, \quad q_{11}=d_rn_4, \quad q_{12}=\rho_2n_7n_5,\\
	&q_{13}=d_nn_5, \quad q_{14}=\rho_3n_7n_6, \quad q_{15}=(d_a+\delta n_4)n_6, \quad q_{16}=\sigma_1 n_5+\sigma_2 n_6, \quad q_{17}=d_in_7,
\end{align*}
and the operators $\varepsilon_i^\pm$ are defined as follows,
\begin{align*}
\varepsilon_i^\pm f(n_1,n_2,n_3,n_4,n_5,n_6,n_7,t)=f(n_1,...,n_i\pm 1,...,n_7,t),\hspace{0.5cm}1\leq i\leq 7.
\end{align*}
If $n_i<0$ for any $1\leq i\leq 7$, then $P(\textbf{n}, t)=0$.

To derive the van Kampen's system size expansion of the master equation~(\ref{master equation}), we rewrite each $n_i(t)$ in the form
\[
n_i(t)=\Omega x_i(t)+\Omega^{1/2}\zeta_i(t),\hspace{0.5cm}1\leq i\leq 7,
\]
where $\Omega x_i(t)=\mathbb{E}[n_i(t)]$, so $\zeta_i(t)$ represent the fluctuations. Replacing the probability density $P(\textbf{n},t)$ by the equivalent probability density $\Pi(\boldsymbol{\zeta},t)$, i.e. with $\Pi(\boldsymbol{\zeta},t)=P(\textbf{n},t)=P\left(\Omega {\bf x}+\Omega^{1/2}{\bf \zeta},t\right)$, the left-hand side of the master equation~(\ref{master equation}) transforms into
\begin{equation}\label{S2}
\displaystyle{\frac{dP(\textbf{n},t)}{dt}=\frac{\partial \Pi}{\partial t}-\sum\limits_{i=1}^{7}\Omega^{1/2}\frac{dx_i}{dt}\frac{\partial \Pi}{\partial \zeta_i}.}
\end{equation}
The operators $\varepsilon_i^{\pm}$ and their product now satisfy the following expansions
\begin{equation}\label{S3}
\begin{array}{l}
\displaystyle{\varepsilon_i^{\pm}=1\pm \Omega^{-1/2}\dfrac{\partial}{\partial \zeta_i}+\frac{1}{2}\Omega^{-1}\frac{\partial^2}{\partial \zeta_i^2}\pm \cdots,}\\\\
\displaystyle{\varepsilon_i^+\varepsilon_j^-=\left(1+\Omega^{-1/2}\frac{\partial}{\partial \zeta_i}+\frac{1}{2}\Omega^{-1}\frac{\partial^2}{\partial \zeta_i^2}+\cdots\right)\left(1- \Omega^{-1/2}\frac{\partial}{\partial \zeta_j}+\frac{1}{2}\Omega^{-1}\frac{\partial^2}{\partial \zeta_j^2}- \cdots\right)}\\
\displaystyle{=1+\Omega^{-1/2}\left(\frac{\partial}{\partial\zeta_i}-\frac{\partial}{\partial\zeta_j}\right)+\Omega^{-1}\left(\frac{1}{2}\dfrac{\partial^2}{\partial\zeta_i^2}-\frac{\partial^2}{\partial\zeta_i\partial\zeta_j}+\frac{1}{2}\dfrac{\partial^2}{\partial\zeta_j^2}\right)+\cdots},\quad 1\leq i,j\leq 7.
\end{array}
\end{equation}
One can easily obtain/introduce $\Omega$-expansions for all parameters $q_i$:
\begin{align*}
q_1&=b_1n_1+b_2n_1^2=b_1\left(\Omega x_1+\Omega^{1/2}\zeta_1\right)+b_2\left(\Omega^2x_1^2+\Omega\zeta_1^2+2\Omega^{3/2}x_1\zeta_1\right)\\
&=b_1\left(\Omega x_1+\Omega^{1/2}\zeta_1\right)+\underbrace{b_2\Omega}_{\widetilde{b}_2}\left(\Omega x_1^2+\zeta_1^2+2\Omega^{1/2}x_1\zeta_1\right)\\
&=\widetilde{b}_2\zeta_1^2+ \left(b_1\zeta_1+2\widetilde{b}_2x_1\zeta_1\right)\Omega^{1/2}+\left(b_1x_1+\widetilde{b}_2x_1^2\right)\Omega,&&
\end{align*}
\begin{equation*}
\resizebox{\textwidth}{!} 
{
$\begin{aligned}
q_2&=d_1n_1+d_2n_1^2+\mu_an_6n_1\\
&=d_1\left(\Omega x_1+\Omega^{1/2}\zeta_1\right)+d_2\left(\Omega^2x_1^2+\Omega\zeta_1^2+2\Omega^{3/2}x_1\zeta_1\right)+\mu_a\left(\Omega x_1+\Omega^{1/2}\zeta_1\right)\left(\Omega x_6+\Omega^{1/2}\zeta_6\right)\\
&=d_1\left(\Omega x_1+\Omega^{1/2}\zeta_1\right)+\underbrace{d_2\Omega}_{\widetilde{d}_2}\left(\Omega x_1^2+\zeta_1^2+2\Omega^{1/2}x_1\zeta_1\right)+\underbrace{\mu_a\Omega}_{\widetilde{\mu}_a}\left(\Omega^{1/2}x_1+\zeta_1\right)\left(\Omega^{1/2} x_6+\zeta_6\right)\\
&=d_1\left(\Omega x_1+\Omega^{1/2}\zeta_1\right)+\widetilde{d}_2\left(\Omega x_1^2+\zeta_1^2+2\Omega^{1/2}x_1\zeta_1\right)+\widetilde{\mu}_a\left(\Omega x_1x_6+\Omega^{1/2}x_1\zeta_6+\Omega^{1/2}x_6\zeta_1+\zeta_1\zeta_6\right)\\
&=\widetilde{\mu}_a\zeta_1\zeta_6+ \left(d_1\zeta_1+2\widetilde{d}_2x_1\zeta_1+\widetilde{\mu}_ax_1\zeta_6+\widetilde{\mu}_ax_6\zeta_1\right)\Omega^{1/2}+\left(d_1x_1+\widetilde{d}_2x_1^2+\widetilde{\mu}_ax_1x_6\right)\Omega,
\end{aligned}$
}
\end{equation*}
\begin{flalign*}
q_3&=\beta n_1n_2=\beta\left(\Omega x_1+\Omega^{1/2}\zeta_1\right)\left(\Omega x_2+\Omega^{1/2}\zeta_2\right)=\underbrace{\beta\Omega}_{\widetilde{\beta}}\left(\Omega^{1/2}x_1+\zeta_1\right)\left(\Omega^{1/2} x_2+\zeta_2\right)\\
&=\widetilde{\beta}\left(\Omega x_1x_2+\Omega^{1/2}x_1\zeta_2+\Omega^{1/2}x_2\zeta_1+\zeta_1\zeta_2\right)= \widetilde{\beta}\zeta_1\zeta_2+ \left(\widetilde{\beta}x_1\zeta_2+\widetilde{\beta}x_2\zeta_1\right)\Omega^{1/2}+ \widetilde{\beta}x_1x_2\Omega,&&
\end{flalign*}
\begin{flalign*}
	q_4=\lambda_{in}=\underbrace{\dfrac{\lambda_{in}}{\Omega}}_{\widetilde{\lambda}_{in}}\Omega=\widetilde{\lambda}_{in}\Omega,&&
\end{flalign*}
\begin{flalign*}
	q_5=d_{in}n_3=d_{in}\left(\Omega x_3+\Omega^{1/2}\zeta_3\right)=d_{in}\zeta_3\Omega^{1/2}+d_{in}x_3\Omega,&&
\end{flalign*}
In a similar way we can easily show
\begin{flalign*}
	q_6=p_1\widetilde{\alpha}\zeta_2\zeta_3+\left(p_1\widetilde{\alpha}x_2\zeta_3+p_1\widetilde{\alpha}x_3\zeta_2\right)\Omega^{1/2}+p_1\widetilde{\alpha}x_2x_3\Omega,&&
\end{flalign*}
\begin{flalign*}
    q_7=p_2\widetilde{\alpha}\zeta_2\zeta_3+\left(p_2\widetilde{\alpha}x_2\zeta_3+p_2\widetilde{\alpha}x_3\zeta_2\right)\Omega^{1/2}+p_2\widetilde{\alpha}x_2x_3\Omega,&&
\end{flalign*}
\begin{flalign*}
    q_8=(1-p_1-p_2)\left[\widetilde{\alpha}\zeta_2\zeta_3+\widetilde{\alpha}x_2\zeta_3+\widetilde{\alpha}x_3\zeta_2\Omega^{1/2}+\widetilde{\alpha}x_2x_3\Omega\right],&&
\end{flalign*}
\begin{flalign*}
	q_9=&\left(\widetilde{\mu}_F\zeta_2\zeta_5+\widetilde{\mu}_a\zeta_2\zeta_6\right)+\left(d_F+\widetilde{\mu}_Fx_2\zeta_5+\widetilde{\mu}_Fx_5\zeta_2+\widetilde{\mu}_ax_2\zeta_6+\widetilde{\mu}_ax_6\zeta_2\right)\Omega^{1/2}\\
	&+\left(d_Fx_2+\widetilde{\mu}_Fx_2x_5+\widetilde{\mu}_ax_2x_6\right)\Omega,&&
\end{flalign*}
\begin{flalign*}
	q_{10}=\widetilde{\rho}_1\zeta_4\zeta_7+\left(\widetilde{\rho}_1x_4\zeta_7+\widetilde{\rho}_1x_7\zeta_4\right)\Omega^{1/2}+\left(\widetilde{\lambda}_r+\widetilde{\rho}_1x_4x_7\right)\Omega,&&
\end{flalign*}
\begin{flalign*}
	q_{11}=d_r\zeta_4\Omega^{1/2}+d_rx_4\Omega,\quad q_{12}=\widetilde{\rho}_2\zeta_5\zeta_7+\left(\widetilde{\rho}_2x_5\zeta_7+\widetilde{\rho}_2x_7\zeta_5\right)\Omega^{1/2}+\widetilde{\rho}_2x_5x_7\Omega,&&
\end{flalign*}
\begin{flalign*}
	q_{13}=d_n\zeta_5\Omega^{1/2}+d_nx_5\Omega,\quad q_{14}=\widetilde{\rho}_3\zeta_6\zeta_7+\left(\widetilde{\rho}_3x_6\zeta_7+\widetilde{\rho}_3x_7\zeta_6\right)\Omega^{1/2}+\widetilde{\rho}_3x_6x_7\Omega,&&
\end{flalign*}
\begin{flalign*}
	q_{15}=\widetilde{\delta}\zeta_4\zeta_6+\left(d_a\zeta_6+\widetilde{\delta}x_4\zeta_6+\widetilde{\delta}x_6\zeta_4\right)\Omega^{1/2}+\left(d_ax_6+\widetilde{\delta}x_4x_6\right)\Omega,&&
\end{flalign*}
\begin{flalign*}
	q_{16}=\left(\sigma_1\zeta_5+\sigma_2\zeta_6\right)\Omega^{1/2}+(\sigma_1x_5+\sigma_2x_6)\Omega,\quad q_{17}=d_i\zeta_7\Omega^{1/2}+d_ix_5\Omega,&&
\end{flalign*}
where
\[
\lambda_{r}=\widetilde{\lambda}_{r}\Omega,\quad \mu_F=\dfrac{\widetilde{\mu}_F}{\Omega}, \quad \alpha=\dfrac{\widetilde{\alpha}}{\Omega}, \quad \delta=\dfrac{\widetilde{\delta}}{\Omega}, \quad \rho_i=\dfrac{\widetilde{\rho}_i}{\Omega}, ~ i=1,2,3.
\]
Substituting expressions (\ref{S2}), (\ref{S3}) and $q_i$'s into the master equation~(\ref{master equation}) shows that the left-hand side of the equation only contains terms of the order $\Omega^{1/2}$ and $\Omega^0$, while the right-hand side has terms of the order $\Omega^{1/2}$, $\Omega^0$, and $\Omega^{-n/2}$, for $n\in \mathbb{N}$. To derive a linear Fokker-Planck equation, we ignore the terms of order $\Omega^{-n/2}$, for $n\in \mathbb{N}$.

Considering the terms of order $\Omega^{1/2}$, i.e. only the terms that are proportional to $\partial \Pi/\partial \zeta_i$, yields
\begin{equation*}
\resizebox{\textwidth}{!} 
{
	$\begin{aligned}
	-\Omega^{1/2}\dfrac{dx_1}{dt}\dfrac{\partial \Pi}{\partial \zeta_1}=&\left(-\Omega^{-1/2}\dfrac{\partial}{\partial \zeta_1}\right)\left[\left(b_1x_1+\widetilde{b}_2x_1^2\right)\Omega\right]\Pi+\left(\Omega^{-1/2}\dfrac{\partial}{\partial \zeta_1}\right)\left[\left(d_1x_1+\widetilde{d}_2x_1^2+\widetilde{\mu}_ax_1x_6\right)\Omega\right]\Pi\\
	&+\left(\Omega^{-1/2}\dfrac{\partial}{\partial\zeta_1}\right)\left[\widetilde{\beta}x_1x_2\Omega\right]\Pi,\\
	-\Omega^{1/2}\dfrac{dx_2}{dt}\dfrac{\partial \Pi}{\partial \zeta_2}=& \left(-\Omega^{-1/2}\dfrac{\partial}{\partial\zeta_2}\right)\left[\widetilde{\beta}x_1x_2\Omega\right]\Pi+\left(\Omega^{-1/2}\dfrac{\partial}{\partial\zeta_2}\right)\left[\left(d_Fx_2+\widetilde{\mu}_Fx_2x_5+\widetilde{\mu}_ax_2x_6\right)\Omega\right]\Pi,\\
	-\Omega^{1/2}\dfrac{dx_3}{dt}\dfrac{\partial \Pi}{\partial \zeta_3}=& \left(-\Omega^{-1/2}\dfrac{\partial}{\partial\zeta_3}\right)\left[\widetilde{\lambda}_{in}\Omega\right]\Pi+\left(\Omega^{-1/2}\dfrac{\partial}{\partial\zeta_3}\right)\left[d_{in}x_3\Omega\right]\Pi+\left(\Omega^{-1/2}\dfrac{\partial}{\partial\zeta_3}\right)\left[p_1\widetilde{\alpha}x_2x_3\Omega\right]\Pi\\
	&\left(\Omega^{-1/2}\dfrac{\partial}{\partial\zeta_3}\right)\left[p_2\widetilde{\alpha}x_2x_3\Omega\right]\Pi+\left(\Omega^{-1/2}\dfrac{\partial}{\partial\zeta_3}\right)\left[(1-p_1-p_2)\widetilde{\alpha}x_2x_3\Omega\right]\Pi,\\
	-\Omega^{1/2}\dfrac{dx_4}{dt}\dfrac{\partial \Pi}{\partial \zeta_4}=& \left(-\Omega^{-1/2}\dfrac{\partial}{\partial\zeta_4}\right)\left[p_1\widetilde{\alpha}x_2x_3\Omega\right]\Pi+\left(-\Omega^{-1/2}\dfrac{\partial}{\partial\zeta_4}\right)\left[\left(\widetilde{\lambda}_r+\widetilde{\rho}_1x_4x_7\right)\Omega\right]\Pi\\
	&+\left(\Omega^{-1/2}\dfrac{\partial}{\partial\zeta_4}\right)\left[d_rx_4\Omega\right]\Pi,\\
	-\Omega^{1/2}\dfrac{dx_5}{dt}\dfrac{\partial \Pi}{\partial \zeta_5}=& \left(-\Omega^{-1/2}\dfrac{\partial}{\partial\zeta_5}\right)\left[p_2\widetilde{\alpha}x_2x_3\Omega\right]\Pi+\left(-\Omega^{-1/2}\dfrac{\partial}{\partial\zeta_5}\right)\left[\widetilde{\rho}_2x_5x_7\Omega\right]\Pi\\
	&-\left(\Omega^{-1/2}\dfrac{\partial}{\partial\zeta_5}\right)\left[d_nx_5\Omega\right]\Pi,\\
	-\Omega^{1/2}\dfrac{dx_6}{dt}\dfrac{\partial \Pi}{\partial \zeta_6}=& \left(-\Omega^{-1/2}\dfrac{\partial}{\partial\zeta_6}\right)\left[(1-p_1-p_2)\widetilde{\alpha}x_2x_3\Omega\right]\Pi+\left(-\Omega^{-1/2}\dfrac{\partial}{\partial\zeta_6}\right)\left[\widetilde{\rho}_3x_6x_7\Omega\right]\Pi\\
	&+\left(\Omega^{-1/2}\dfrac{\partial}{\partial\zeta_6}\right)\left[\left(d_ax_6+\widetilde{\delta}x_4x_6\right)\Omega\right]\Pi,\\
	-\Omega^{1/2}\dfrac{dx_7}{dt}\dfrac{\partial \Pi}{\partial \zeta_7}=& \left(-\Omega^{-1/2}\dfrac{\partial}{\partial\zeta_7}\right)\left[(\sigma_1x_5+\sigma_2x_6)\Omega\right]\Pi+\left(\Omega^{-1/2}\dfrac{\partial}{\partial\zeta_7}\right)\left[d_ix_5\Omega\right]\Pi.
	\end{aligned}$
}
\end{equation*}
After simplification, we obtain the following system of equations that describes macroscopic behaviour of the model
\begin{align}\label{secondmodel_sup}
\begin{split}
\dfrac{dx_1}{dt}&=b_1x_1+\widetilde{b}_2x_1^2-d_1x_1-\widetilde{d}_2x_1^2-\widetilde{\beta}x_1x_2-\widetilde{\mu}_a x_1x_6,\\
\dfrac{dx_2}{dt}&=\widetilde{\beta}x_1x_2-d_Fx_2-\widetilde{\mu}_Fx_2x_5-\widetilde{\mu}_a x_2x_6,\\
\dfrac{dx_3}{dt}&=\widetilde{\lambda}_{in}-d_{in}x_3-\widetilde{\alpha}x_2x_3,\\
\dfrac{dx_4}{dt}&=\widetilde{\lambda}_r-d_rx_4+p_1\widetilde{\alpha}x_2x_3+\widetilde{\rho}_1x_4x_7,\\
\dfrac{dx_5}{dt}&=p_2\widetilde{\alpha}x_2x_3-d_nx_5+\widetilde{\rho}_2 x_5x_7,\\
\dfrac{dx_6}{dt}&=(1-p_1-p_2)\widetilde{\alpha}x_2x_3-d_ax_6-\widetilde{\delta}x_4x_6+\widetilde{\rho}_3x_6x_7,\\
\dfrac{dx_7}{dt}&=\sigma_1x_5+\sigma_2x_6-d_ix_7.
\end{split}
\end{align}
Terms of order $\Omega^0$ give the following Fokker-Planck equation
\begin{equation*}
\resizebox{\textwidth}{!} 
{
	$\begin{aligned}
	\dfrac{\partial\Pi}{\partial t}=-\bigg[&\left(b_1+2\widetilde{b}_2x_1-d_1-2\widetilde{d}_2x_1-\widetilde{\mu}_ax_6-\widetilde{\beta}x_2\right)\dfrac{\partial(\zeta_1\Pi)}{\partial\zeta_1}-\widetilde{\beta}x_1\dfrac{\partial(\zeta_2\Pi)}{\partial\zeta_1}-\widetilde{\mu}_ax_1\dfrac{\partial(\zeta_6\Pi)}{\partial\zeta_1}\\
	&+\widetilde{\beta}x_2\dfrac{\partial(\zeta_1\Pi)}{\partial\zeta_2}+\left(\widetilde{\beta}x_1-d_F-\widetilde{\mu}_Fx_5-\widetilde{\mu}_ax_6\right)\dfrac{\partial(\zeta_2\Pi)}{\partial\zeta_2}-\widetilde{\mu}_Fx_2\dfrac{\partial(\zeta_5\Pi)}{\partial\zeta_2}-\widetilde{\mu}_ax_2\dfrac{\partial(\zeta_6\Pi)}{\partial\zeta_2}\\
	&-\widetilde{\alpha}x_3\dfrac{\partial(\zeta_2\Pi)}{\partial\zeta_3}-\left(d_{in}+\widetilde{\alpha}x_2\right)\dfrac{\partial(\zeta_3\Pi)}{\partial\zeta_3}+p_1\widetilde{\alpha}x_3\dfrac{\partial(\zeta_2\Pi)}{\partial\zeta_4}+p_1\widetilde{\alpha}x_2\dfrac{\partial(\zeta_3\Pi)}{\partial\zeta_4}+\left(\widetilde{\rho}_1x_7-d_r\right)\dfrac{\partial(\zeta_4\Pi)}{\partial\zeta_4}\\
	&+\widetilde{\rho}_1x_4\dfrac{\partial(\zeta_7\Pi)}{\partial\zeta_4}+p_2\widetilde{\alpha}x_3\dfrac{\partial(\zeta_2\Pi)}{\partial\zeta_5}+p_2\widetilde{\alpha}x_2\dfrac{\partial(\zeta_3\Pi)}{\partial\zeta_5}+\left(\widetilde{\rho}_2x_7-d_n\right)\dfrac{\partial(\zeta_5\Pi)}{\partial\zeta_5}+\widetilde{\rho}_2x_5\dfrac{\partial(\zeta_7\Pi)}{\partial\zeta_5}\\
	&+(1-p_1-p_2)\widetilde{\alpha}x_3\dfrac{\partial(\zeta_2\Pi)}{\partial\zeta_6}+(1-p_1-p_2)\widetilde{\alpha}x_2\dfrac{\partial(\zeta_3\Pi)}{\partial\zeta_6}-\widetilde{\delta}x_6\dfrac{\partial(\zeta_4\Pi)}{\partial\zeta_6}\\
	&+\left(\widetilde{\rho}_3x_7-d_a-\widetilde{\delta}\right)x_4\dfrac{\partial(\zeta_6\Pi)}{\partial\zeta_6}+\widetilde{\rho}_3x_6\dfrac{\partial(\zeta_7\Pi)}{\partial\zeta_6}+\sigma_1\dfrac{\partial(\zeta_5\Pi)}{\partial\zeta_7}+\sigma_2\dfrac{\partial(\zeta_6\Pi)}{\partial\zeta_7}-d_i\dfrac{\partial(\zeta_7\Pi)}{\partial\zeta_7}\bigg]\\
	+\dfrac{1}{2}&\bigg\{\left(b_1x_1+\widetilde{b}_2x_1^2+d_1x_1+\widetilde{d}_2x_1^2+\widetilde{\beta}x_1x_2+\widetilde{\mu}_ax_1x_6\right)\dfrac{\partial^2\Pi}{\partial\zeta_1^2}-2\widetilde{\beta}x_1x_2\dfrac{\partial^2\Pi}{\partial\zeta_1\partial\zeta_2}\\
	&+\left(\widetilde{\beta}x_1x_2+d_Fx_2+\widetilde{\mu}_Fx_2x_5+\widetilde{\mu}_ax_2x_6\right)\dfrac{\partial^2\Pi}{\partial\zeta_2^2}+\left(\widetilde{\lambda}_{in}+d_{in}x_3+\widetilde{\alpha}x_2x_3\right)\dfrac{\partial^2\Pi}{\partial\zeta_3^2}\\
	&-2p_1\widetilde{\alpha}x_2x_3\dfrac{\partial^2\Pi}{\partial\zeta_3\partial\zeta_4}-2p_2\widetilde{\alpha}x_2x_3\dfrac{\partial^2\Pi}{\partial\zeta_3\partial\zeta_5}-2(1-p_1-p_2)\widetilde{\alpha}x_2x_3\dfrac{\partial^2\Pi}{\partial\zeta_3\partial\zeta_6}\\
	&+\left(\widetilde{\lambda}_r+d_rx_4+p_1\widetilde{\alpha}x_2x_3+\widetilde{\rho}_1x_4x_7\right)\dfrac{\partial^2\Pi}{\partial\zeta_4^2}+\left(p_2\widetilde{\alpha}x_2x_3+d_nx_5+\widetilde{\rho}_2 x_5x_7\right)\dfrac{\partial^2\Pi}{\partial\zeta_5^2}\\
	&+\left[(1-p_1-p_2)\widetilde{\alpha}x_2x_3+d_ax_6+\widetilde{\delta}x_4x_6+\widetilde{\rho}_3x_6x_7\right]\dfrac{\partial^2\Pi}{\partial\zeta_6^2}+\left(\sigma_1x_5+\sigma_2x_6+d_ix_7\right)\dfrac{\partial^2\Pi}{\partial\zeta_7^2}\bigg\}.
	\end{aligned}$
}
\end{equation*}
This equation can be equivalently rewritten in the form
\begin{equation*}
\dfrac{\partial\Pi(\boldsymbol{\zeta},t)}{\partial t}=-\sum\limits_{i,j}A_{ij}\dfrac{\partial}{\partial \zeta_i}(\zeta_j\Pi)+\dfrac{1}{2}\sum\limits_{i,j}B_{ij}\dfrac{\partial^2\Pi}{\partial\zeta_i \partial\zeta_j},
\end{equation*}
where $A$ is the Jacobian matrix of system~(\ref{secondmodel_sup})
\[
\resizebox{\linewidth}{!}{%
	$\displaystyle
	A=
	\begin{pmatrix}
	b_1+2\widetilde{b}_2x_1-d_1-2\widetilde{d}_2x_1-\widetilde{\mu}_ax_6-\widetilde{\beta}x_2 & -\widetilde{\beta}x_1 & 0 & 0 & 0 & -\widetilde{\mu}_ax_1 & 0 \\
	\widetilde{\beta}x_2 & \widetilde{\beta}x_1-d_F-\widetilde{\mu}_Fx_5-\widetilde{\mu}_ax_6 & 0 & 0 & -\widetilde{\mu}_Fx_2 & -\widetilde{\mu}_ax_2 & 0 \\
	0 & -\widetilde{\alpha}x_3 & -d_{in}-\widetilde{\alpha}x_2 & 0 & 0 & 0 & 0 \\
	0 & p_1\widetilde{\alpha}x_3 & p_1\widetilde{\alpha}x_2 & \widetilde{\rho}_1x_7-d_r & 0 & 0 & \widetilde{\rho}_1x_4\\
	0 & p_2\widetilde{\alpha}x_3 & p_2\widetilde{\alpha}x_2 & 0 & \widetilde{\rho}_2x_7-d_n & 0 & \widetilde{\rho}_2x_5\\
	0 & (1-p_1-p_2)\widetilde{\alpha}x_3 & (1-p_1-p_2)\widetilde{\alpha}x_2 & -\widetilde{\delta}x_6 & 0 & \widetilde{\rho}_3x_7-d_a-\widetilde{\delta}x_4 & \widetilde{\rho}_3x_6\\
	0 & 0 & 0 & 0 & \sigma_1 & \sigma_2 & -d_i 
	\end{pmatrix}
	$}
\]
$B$ is a $7\times 7$ symmetric matrix given by
\begin{center}
	$B_{ij}$ = $\begin{cases} b_1x_1+\widetilde{b}_2x_1^2+d_1x_1+\widetilde{d}_2x_1^2+\widetilde{\beta}x_1x_2+\widetilde{\mu}_a x_1x_6, &\mbox{ if } (i,j) = (1,1),\\ 
	\widetilde{\beta}x_1x_2+d_Fx_2+\widetilde{\mu}_Fx_2x_5+\widetilde{\mu}_a x_2x_6, & \mbox{ if } (i,j) = (2,2),\\
	\widetilde{\lambda}_{in}+d_{in}x_3+\widetilde{\alpha}x_2x_3, & \mbox{ if } (i,j) = (3,3),\\
	\widetilde{\lambda}_r+d_rx_4+p_1\widetilde{\alpha}x_2x_3+\widetilde{\rho}_1x_4x_7, & \mbox{ if } (i,j) = (4,4),\\
	p_2\widetilde{\alpha}x_2x_3+d_nx_5+\widetilde{\rho}_2 x_5x_7, & \mbox{ if } (i,j) = (5,5),\\
	(1-p_1-p_2)\widetilde{\alpha}x_2x_3+d_ax_6+\widetilde{\delta}x_4x_6+\widetilde{\rho}_3x_6x_7, & \mbox{ if } (i,j) = (6,6),\\
	\sigma_1x_5+\sigma_2x_6+d_ix_7, & \mbox{ if } (i,j) = (7,7),\\
	-\widetilde{\beta}x_1x_2, & \mbox{ if } (i,j) = (1,2)\mbox{ or }(2,1),\\
	-p_1\widetilde{\alpha}x_2x_3, & \mbox{ if } (i,j) = (3,4)\mbox{ or }(4,3),\\
	-p_2\widetilde{\alpha}x_2x_3, & \mbox{ if } (i,j) = (3,5)\mbox{ or }(5,3),\\
	-(1-p_1-p_2)\widetilde{\alpha}x_2x_3, & \mbox{ if } (i,j) = (3,6)\mbox{ or }(6,3),\\
	0, & \mbox{ otherwise.}
	\end{cases}$
\end{center}

\end{document}